\documentclass[12pt]{article}
\usepackage{amsthm}
\usepackage{amsmath}
\usepackage{natbib}
\usepackage[colorlinks,citecolor=blue,urlcolor=blue,filecolor=blue,backref=page]{hyperref}
\usepackage{graphicx}

\usepackage{booktabs}

\usepackage{enumerate}
\usepackage{url} 

\usepackage[export]{adjustbox} 
\usepackage[group-separator={,}]{siunitx}

\usepackage[utf8]{inputenc}

\usepackage{caption}
\usepackage{subcaption}

\usepackage{amsfonts}
\usepackage{amssymb}
\usepackage{adjustbox}
\usepackage{hhline}

\usepackage{enumitem}
\usepackage{bm}

\usepackage[boxruled]{algorithm2e}

\usepackage{algpseudocode}
\usepackage{physics}

\usepackage{mathrsfs}

\usepackage{mathtools}

\usepackage{tikz}
\usepackage{pgfplots}
\pgfplotsset{compat=newest}
\usetikzlibrary{shapes.geometric,arrows,fit,matrix,positioning}
\tikzset
{
    treenode/.style = {ellipse, draw=white, anchor=north},
    terminal/.style = {rectangle, draw=white, minimum width=0.6cm, anchor=north},
}
\usepackage{forest}

\newtheorem{theorem}{Theorem}
\newtheorem{corollary}{Corollary}[theorem]

\newtheorem{definition}{Definition}
\newtheorem{example}{Example}

\newcommand{\x}{\mathbf{x}}
\newcommand{\xx}{\mathbf{x}}  
\newcommand{\X}{\mathbf{X}}

\newcommand{\z}{\mathbf{z}}

\newcommand{\uu}{\mathbf{u}}

\newcommand{\bfW}{\mathbf{W}}

\newcommand{\Cov}{\text{Cov}}
\newcommand{\corr}{\text{corr}}
\newcommand{\Var}{\text{Var}}

\newcommand{\E}{E}

\newcommand{\I}{\mathbf{I}}

\newcommand{\T}{\mathcal{T}}

\newcommand{\iset}{\mathcal{X}}

\newcommand{\bM}{\mathbf{M}}

\newcommand{\ind}{\stackrel{ind}{\sim}}

\makeatletter
\DeclareFontEncoding{LS1}{}{}
\DeclareFontSubstitution{LS1}{stix}{m}{n}
\DeclareMathAlphabet{\mathscr}{LS1}{stixscr}{m}{n}
\makeatother

\newcommand{\Normal}{\text{N}}

\newcommand{\PG}{P\'olya-Gamma }

\newcommand{\EE}{E}
\newcommand{\ep}{\varepsilon}
\newcommand{\prior}{treeSB}

\newcommand*\dif{\mathop{}\!\mathrm{d}}

\newcommand{\bep}{\bm{\ep}}

\newcommand{\bgamma}{\bm{\gamma}}
\newcommand{\bkappa}{\bm{\kappa}}
\newcommand{\bpsi}{\bm{\psi}}
\newcommand{\fe}{\gamma}
\newcommand{\res}{u}
\newcommand{\rej}{v}

\newcommand{\bfe}{\bm{\fe}}

\newcommand{\rep}{\phi}
\newcommand{\Rep}{\Phi}
\newcommand{\bRep}{\bm{\Rep}}

\newcommand{\bSigma}{\bm{\Sigma}}
\newcommand{\bmu}{\bm{\mu}}
\newcommand{\bOmega}{\bm{\Omega}}

\newcommand{\cJ}{\mathcal{J}}
\newcommand{\cS}{\mathcal{S}}
\newcommand{\cD}{\mathcal{D}}
\newcommand{\cL}{\mathcal{L}}

\usepackage{stackengine}
\usepackage{scalerel}
\def\rlwd{.4pt}
\def\rlht{1.1pt}
\def\shatvrule{\rule{\rlwd}{\rlht}}
\def\shat#1{%
 \ThisStyle{%
  \setbox0=\hbox{$\SavedStyle#1$}%
  \stackon[0pt]{\stackon[1pt]{\ensuremath{\SavedStyle#1}}{%
    \shatvrule\kern\wd0\kern-\rlwd\kern-\rlwd\shatvrule}}%
    {\rule{\wd0}{\rlwd}}%
 }%
}

\usepackage{tikz-cd}
\usepackage{array,multirow}

\usepackage{authblk}

\begin{document}

\def\spacingset#1{\renewcommand{\baselinestretch}%
{#1}\small\normalsize} \spacingset{1}


\date{}
\title{\bf A tree perspective on stick-breaking models in covariate-dependent mixtures}
    \author[1]{Akira Horiguchi}
  \author[2,1]{Cliburn Chan}
  \author[1,2]{Li Ma\thanks{Send correspondence to \texttt{li.ma@duke.edu}}}
  \affil[1]{Department of Statistical Science, Duke University \protect\\ 
  214 Old Chemistry, Box 90251, Durham, NC 27708}
  \affil[2]{Department of Biostatistics and Bioinformatics, Duke University \protect\\ 
  2424 Erwin Road, Suite 1102, Hock Plaza Box 2721, Durham, NC 27710}
\maketitle

\bigskip
\begin{abstract}
Stick-breaking (SB) processes are often adopted in Bayesian  mixture models for generating mixing weights.
When covariates influence the sizes of clusters, SB mixtures are particularly convenient as they can leverage their connection to binary regression to ease both the specification of covariate effects and posterior computation. Existing SB models are typically constructed based on continually breaking a single remaining piece of the unit stick. We view this from a dyadic tree perspective in terms of a lopsided bifurcating tree that extends only in one side. We show that two unsavory characteristics of SB models are in fact largely due to this lopsided tree structure. We consider a generalized class of SB models with alternative bifurcating tree structures and examine the influence of the underlying tree topology on the resulting Bayesian analysis in terms of prior assumptions, posterior uncertainty, and computational effectiveness. In particular, we provide evidence that a balanced tree topology, which corresponds to continually breaking all remaining pieces of the unit stick, can resolve or mitigate these undesirable properties of SB models that rely on a lopsided tree.
\end{abstract}

\noindent%
{\it Keywords:}  Discrete random measure, Bayesian nonparametric, tail-free process, clustering analysis, flow cytometry.
\vfill

\newpage

\section{Introduction}
\label{sec:intro}

Mixture models are a popular approach to clustering analysis. A mixture model's weights are often generated by stick-breaking processes \citep{Sethuraman1994,Ishwaran2001} largely due to their tractable posterior computation. Stick-breaking processes are particularly well-suited when covariates influence cluster sizes; these models can readily incorporate covariate influence through binary regression at each stick break and can leverage efficient posterior computation strategies developed for binary regression models.
A notable example is \cite{Rodriguez2011}'s probit stick-breaking whose Markov chain Monte Carlo algorithm relies on \cite{albert1993bayesian}'s truncated Gaussian data augmentation for probit regression. 
Similarly, for logit stick-breaking \cite{Rigon2021} provide various posterior sampling methods that rely on \cite{Polson2013}'s Pólya-Gamma data augmentation for logistic regression.
\cite{linderman2015advances} also employ Pólya-Gamma data augmentation for stick-breaking for dependent multinomial modeling. 
Alternatively, mixing-weight dependence could be induced by partially exchangeable observations \citep{de1938condition,diaconis1980finetti,diaconis1988recent} or by splitting variables without explicitly incorporating covariates \citep[e.g.][]{fuentes2010new,favaro2012stick,favaro2016stick,gil2020beta,gil2023stick}. 
Our paper will focus on dependence that is explicitly characterized by covariates 
in terms of their effects on the mixing weights.

While our main applied motivation is the need for covariate-dependent clustering in the analysis of flow cytometry data (see Section~\ref{sec:posterioruncertainty}),
stick-breaking models and their dependent variants have also been adopted in a wide range of other applied contexts. Such contexts include time-series data \citep{griffin2011stick,bassetti2014beta}, spatial data \citep{griffin2006order,duan2007generalized,reich2007multivariate,dunson2008kernel}, and space-time data \citep{hossain2013space,grazian2024spatio} among many other examples. Deepening the understanding of this useful model class and improving its design and construction can thus lead to better statistical practice across these applications.

Although all discrete random measures in principle admit a stick-breaking representation, identifying the pairing between their stick-breaking construction and other constructive representations is an interesting ongoing endeavor in the Bayesian nonparametrics literature. Following the canonical construction of stick-breaking representations for the Dirichlet process \citep{Ferguson1973,Sethuraman1994} and the Pitman-Yor process \citep{Pitman1997}, the stick-breaking representation has been successfully identified for the broad class of homogeneous normalized random measures with independent increments  \citep{regazzini2003distributional,favaro2016stick}. This class contains as special cases several processes whose stick-breaking representations were previously identified, including the normalized generalized gamma process \citep{lijoi2007controlling}, which in turn contains the normalized inverse Gaussian process \citep{lijoi2005hierarchical,favaro2012stick} and the Dirichlet process.
Beyond random probability measures, stick-breaking constructions have also been identified for some completely random measures such as the beta process \citep{teh2007stick,paisley2010stick,paisley2012stick,broderick2012beta,hjort1990nonparametric}, 
which has been applied in latent factor models \citep{ghahramani2005infinite,thibaux2007hierarchical} and mixed-membership models \citep{fox2014mixed}.

While much of the existing literature investigates stick-breaking representations for known random measures, the reverse direction of identifying alternative representations for a given stick-breaking process is also of interest. Such a representation, either in the form of a random measure or otherwise, would provide both theoretical and practical insights into key properties of the stick-breaking model. The only representations for stick-breaking processes with independent splitting variables so far identified are the Dirichlet and Pitman-Yor processes \citep{pitman1996random,favaro2012stick}. As a reviewer kindly pointed out, this appears to be due to the fact that the Dirichlet and the Pitman-Yor processes are the only stick-breaking processes with independent weights being invariant under size-biased permutations as proved in \cite{pitman1996random}, and without such a property it seems impossible with current techniques to derive alternative representations for stick-breaking models with independent splitting variables. 

\begin{figure}
    \centering
    \includegraphics[width=\textwidth]{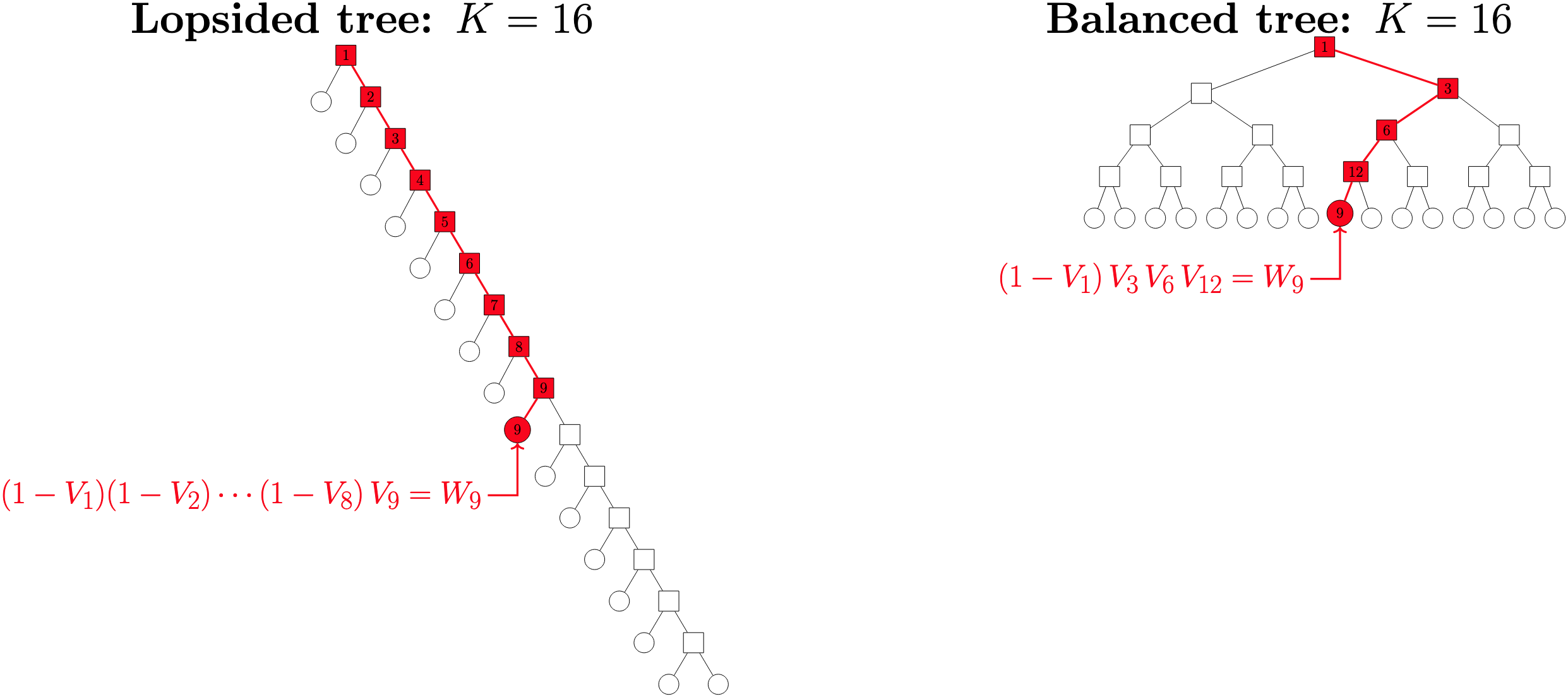}
    \caption{Binary tree representation of stick-breaking schemes. Each scheme's initial stick break is represented as a root node; each subsequent stick break is represented by an internal node. The mass of a stick-break's ``left (right) piece'' is sent to its left (right) child. 
    The $K$ leaf nodes represent the scheme's $K$ weights.
    As an example, weight $W_9$ is a product of nine stick breaks in the lopsided-tree scheme but only four stick breaks in the balanced-tree scheme.
    }
    \label{fig:treestructure}
\end{figure}

Existing stick-breaking models for random probability measures generate a set of unit-sum weights through continually breaking pieces off a unit stick, one piece at a time. 
This mechanism can be viewed through a dyadic tree perspective in which each stick breaking step corresponds to a bifurcating split of the remaining stick resulting in two ``children'', one representing a leaf (i.e., terminal) node corresponding to the piece broken off from the stick, and the other a non-leaf (i.e., interior) node corresponding to the remaining stick to be further broken.  
Without loss of generality, we use the {\em left} child of each split to represent the leaf node (i.e., the piece broken off) and the {\em right} child as the remaining stick. Following this delineation, the tree has a lopsided shape which extends only in the right side as shown in Figure~\ref{fig:treestructure}. Such a stick-breaking mechanism, which we shall refer to as a lopsided-tree stick-breaking, has been widely applied in the Bayesian nonparametric literature in constructing discrete random measures.

Our work is motivated by the observation that some well-known properties of existing stick-breaking models, for example the stochastic ordering of the resulting weights, are in fact attributable to this lopsided-tree structure, whereas the key benefits of stick-breaking models, such as the ability to transform modeling and Bayesian computation into binary regression problems, require only a dyadic tree structure which need not be lopsided. It is therefore a promising strategy to resolve some limitations of stick-breaking models by generalizing beyond the lopsided tree to different dyadic tree structures, while maintaining the modeling and computational benefits. 
A natural question is whether such a generalization is practically useful. 
The short answer, as we shall demonstrate, is indeed yes. The choice of the tree structure can substantially influence statistical inference under stick-breaking models.

More specifically, we show that in the context of covariate-dependent mixture modeling, which is a major field of application for stick-breaking models, common model specifications with a lopsided tree structure induce two undesirable characteristics that can severely deteriorate the quality of Bayesian inference in terms of increased posterior uncertainty and reduced computational efficiency. Some of these properties have been noted previously in the literature, but they were never attributed to the underlying lopsided tree structure, which we show is a major culprit.  
In addition, we show through a combination of theoretical analysis, numerical experiments, and a case study that these two undesirable features of stick-breaking models can be resolved by simply adopting an alternative dyadic tree structure, 
namely a ``balanced'' tree (illustrated in Figure~\ref{fig:treestructure}), corresponding to continually breaking off {\em both} remaining sticks at each stick-breaking step, without complicating the modeling and computational recipes that stick-breaking models enjoy.

What are these two undesirable characteristics of (lopsided-tree) stick-breaking processes  in covariate-dependent mixture modeling? The first characteristic, which will be the focus of Section~\ref{sec:priorcorrelation}, is that under commonly adopted ``default'' specifications, covariate-dependent stick-breaking models 
can induce a strong positive prior correlation in the random measures over covariate values, which can cause excessive smoothing 
across the covariates even when the corresponding cluster sizes vary sharply across covariate values, thereby deprecating the clustering at each covariate value.
This phenomenon was first explained in \cite{Rodriguez2011} in the context of probit-stick-breaking models as 
``a consequence of our use of a common set of atoms at every [covariate]; even if the set of weights are independent from each other, the fact that the atoms are shared means that the distributions cannot be independent.''
Interestingly, we show that it is the lopsided-tree structure underlying standard stick-breaking that multiplies the effects of shared atoms on prior correlation.

The second characteristic, which will be the theme of Section~\ref{sec:posterioruncertainty}, concerns the precision of the inference in terms of the posterior uncertainty of covariate effects on mixing weights. 
Common prior specifications of covariate effects in stick-breaking weights introduce many competing mechanisms in a mixture model. 
Such specifications include stochastic ordering of the weights and the large number of stick breaks the weights is a product of (on average), which both introduce mechanisms that, at best, add unnecessary layers of complexity and, at worst, actively degrade the quality of posterior inference. 

To finish the introduction, we relate to some relevant papers in the literature that involve tree-structured Bayesian nonparametric models. 
\cite{Ghahramani2010} model hierarchical data by interleaving two lopsided stick-breaking processes to allow a wide range of latent tree structures (e.g. nodes with any number of children) to be inferred.
On the other hand, stick-breaking models over a general tree structure are closely related to tree-structured random measures such as the P\'olya tree, and their generalizations involving covariate-dependence has a parallel development for tree-structured random measures, namely the covariate-dependent tail-free model introduced by \cite{Jara2011}.
To our knowledge, such processes have not been applied as a \textit{discrete} prior for mixing distributions except in \cite{Cipolli2017} who propose the finite P\'olya tree as a prior for the mixing distributions but do not argue the use of their approach over lopsided stick-breaking processes.
\cite{Stefanucci2021} introduce another weight-generating mechanism along a balanced bifurcating tree that includes stick-breaking models as special cases, but this generalization assigns weights to every node of the tree, which makes this approach lose the main computational benefits of stick-breaking models in incorporating covariates.
Finally, \cite{Ren2011} note a covariate-dependent stick-breaking process ``may be viewed as a mixture-of-experts model'' \citep{jordan1994hierarchical,peng1996bayesian,bishop2002bayesian} which uses a binary tree to define a covariate-dependent mixture distribution. 
Although we investigate the influence of the tree structure mainly in the context of stick-breaking models, the lessons drawn are also relevant for mixture-of-experts models given the connection between the two model classes.

\section{Stick-breaking models from a tree perspective}

We start by introducing our tree-based stick-breaking construction in the absence of covariates and then extend the construction to include covariates. This broader class of models contains existing stick-breaking models as a special case corresponding to a particular lopsided tree structure. This will let us later examine the impact of the tree structure on Bayesian inference through the lens of comparing models within this broader class.

This paper assumes conditionally independent and identically distributed (i.i.d.) observations $y_1, \ldots, y_n \in \mathbb{R}^d$, given a mixing measure $G$, from a sampling density of the mixture form $f(\cdot; G) = \int_{\Theta} h(\cdot; \theta) \dif G(\theta)$,
where $h\colon \mathbb{R}^d \times \Theta \rightarrow \mathbb{R}^+$ is a parametric mixing kernel density, e.g., Gaussian density $h(y; \theta) = (2\pi \sigma^2)^{-1/2} \exp\left\{-(y-\mu)^2/(2\sigma^2)\right\}$ with $\theta=(\mu, \sigma^2) \in \mathbb{R} \times \mathbb{R}^+$.

If a mixing measure $G$ on $\Theta$ is equipped with a stick-breaking prior, each realization of $G$ is, with probability one, a discrete measure $G = \sum_{k=1}^{\infty} W_k \delta_{\theta_k}$
where the atoms $\theta_k$ are generated independently from a probability measure $G_0$ on $\Theta$ and the nonnegative weights $W_k$, which are generated by breaking a unit stick, are independent of the atoms $\theta_k$.
This turns the mixture density $\int_{\Theta} h(\cdot; \theta) \dif G(\theta)$ into the discrete sum $\sum_{k=1}^{\infty} W_k h(\cdot; \theta_k)$.

The remainder of this section examines the traditional stick-breaking scheme from a tree perspective, and then introduces a tree-based generalization to stick-breaking priors.

\subsection{Stick-breaking models without covariates}
\label{sec:tree stick-breaking}

A weight-generation process aims to produce nonnegative quantities that sum to unity.
A stick-breaking process achieves this by successively breaking off pieces of a stick of initially unit length,
where the lengths of the resulting pieces become the desired weights. 
In traditional stick-breaking, each stick break produces a piece that is untouched afterwards while the other piece---the ``remaining stick''---breaks again and again.
More formally, to begin, a unit-length stick breaks at location $V_1 \in [0,1]$ so that the piece that breaks off has length $V_1$.
The remaining stick, of length $1-V_1$, then breaks at (relative) location $V_2 \in [0,1]$ so that the new piece that breaks off has length $(1-V_1)V_2$.
The remaining stick, now of length $(1-V_1)(1-V_2)$, then breaks at location $V_3 \in [0,1]$ so that the new piece that breaks off has length $(1-V_1)(1-V_2)V_3$, so on and so forth.
For each $k\in\mathbb{N}$, the $k^{\text{th}}$ stick break at location $V_k \in [0,1]$ produces a piece that breaks off with length $V_k \prod_{l=1}^{k-1} (1-V_l)$, which sets the value of weight $W_k$; the remaining stick then breaks at location $V_{k+1} \in [0,1]$.
The process can either continue infinitely or terminate after a preset finite number of breaks, in which case the final remaining stick sets the last weight.

A stick-breaking scheme can be identified with a bifurcating tree whose nodes correspond to the pieces that arise during the stick breaking procedure.
We construct the binary tree by first assigning the initial unit-length stick to the tree's root node, and iteratively apply the following steps.
If a piece of the stick does not break further, the piece corresponds to a leaf node, i.e., a node with no children.
If a piece is further broken instead, the two resulting pieces correspond to two children nodes. 
The traditional stick-breaking scheme's identified bifurcating tree 
(see the left of Figure~\ref{fig:treestructure})
is the most ``lopsided'' as none of its nodes have children that both divide further. 
It is the deepest tree possible for generating a given number of pieces. 
We shall thus refer to the traditional stick-breaking strategy as \textit{lopsided-tree stick-breaking}.

Given this tree view of stick-breaking, it is natural to consider stick-breaking schemes corresponding to binary tree topologies that differ from the lopsided one. 
To this end, it will be convenient to index the root node by the empty string $\emptyset$ and each other tree node $\bep$ by a finite string of 0s and 1s,
where each digit indicates whether a node along the path from root to $\bep$ is a left (0) or right (1) child of its parent.
In general, any node at level $m$ of the tree (the root node is at level 0, its two children are at level 1, and so on) is indexed by a $m$-length binary string $\ep_1 \ep_2 \cdots \ep_m$, where the string $\bep \bep'$ denotes the concatenation of finite strings $\bep$ and $\bep'$.
The set of all finite binary strings (including $\emptyset$) is denoted by $\EE^* \coloneqq \cup_{m=0}^{\infty} \{0,1\}^m$.
This machinery allows us to formally introduce stick-breaking strategies based on general tree topologies. 

We give particular attention to the stick-breaking scheme corresponding to a ``balanced'' bifurcating tree, in which all nodes (up to a maximum level) are split into two children as illustrated at 
the right of Figure~\ref{fig:treestructure}. 
We refer to this stick-breaking scheme as a \textit{balanced-tree stick-breaking}. 
Opposite of the lopsided-tree scheme, the balanced-tree scheme results in the most shallow tree structure needed to generate any given number of weights.
In this scheme, each stick break produces two pieces that \emph{both} break again until the total number of breaks exceeds a preset threshold (we argue against allowing infinitely many breaks in Section \ref{sec:choiceofnumberofleaves}).
More formally, the unit-length stick $I_{\emptyset}$ breaks according to $V_{\emptyset}$ so that the left piece $I_0$ has length $|I_0| = |I_{\emptyset}|V_{\emptyset} = V_{\emptyset}$ and the right piece has length $|I_1| = |I_{\emptyset}|(1-V_{\emptyset})=1-V_{\emptyset}$.
These two pieces $I_0$ and $I_1$ then break similarly according to respective splitting variables $V_0$ and $V_1$ so that the resulting four pieces $I_{00}$, $I_{01}$, $I_{10}$, and $I_{11}$ have lengths
$|I_{00}| = V_{\emptyset} V_0$, $|I_{01}| = V_{\emptyset} (1-V_0)$, $|I_{10}| = (1-V_{\emptyset}) V_1$, and $|I_{11}| = (1-V_{\emptyset}) (1-V_1)$.
These four pieces then break similarly according to respective splitting variables $V_{00}$, $V_{01}$, $V_{10}$, and $V_{11}$ to produce the eight pieces $I_{000}$, $I_{001}$, $I_{010}$, $I_{011}$, $I_{100}$, $I_{101}$, $I_{110}$, and $I_{111}$ with lengths defined similarly.
If each piece is only allowed to be a product of at most $m \in \mathbb{N}$ stick breaks, this recursive procedure produces $2^m$ leaf nodes each with a stick piece.
The lengths of these pieces define the desired weights: $W_{\bep} \coloneqq |I_{\bep}|$ for each $\bep \in \{0,1\}^m$.

The value of any stick-breaking weight at a node $\bep = \ep_1 \cdots \ep_m$ can be expressed as
\begin{equation}
\label{eq:genweights}
W_{\ep_1 \cdots \ep_m} = \prod_{l=1}^m V_{\ep_1 \cdots \ep_{l-1}}^{1 - \ep_l} (1-V_{\ep_1 \cdots \ep_{l-1}})^{\ep_l}
\quad \quad  V_{\bep} \sim F_{\bep} \text { for all } \bep \in E^*
\end{equation}
where by convention $\ep_1 \cdots \ep_{l-1} = \emptyset$ if $l=1$,
and the splitting variables are distributed according to a countable sequence $\{F_{\bep} \colon \bep \in E^*\}$ of distributions each with full support on $[0,1]$
and the tail-free condition $V_{\emptyset} \perp \{V_0, V_1\} \perp \{V_{00}, V_{01}, V_{10}, V_{11}\} \perp \cdots$ \citep{Freedman1963,Fabius1964}.

We can now define the \textit{tree stick-breaking} class of priors, which includes traditional stick-breaking (where each level of the tree other than level zero has exactly one weight and that node takes the form $1 \cdots 10$) and balanced-tree stick-breaking but also admits more general tree structures.
\begin{definition}
    If $G_0$ is a probability measure, $\{F_{\bep}: \bep \in E^*\}$ is a sequence of distributions each with support $[0,1]$, and $\tau$ is a binary tree structure, 
    we say a probability measure $G$ is equipped with a tree stick-breaking prior with parameters $G_0$, $\{F_{\bep}\}$, and $\tau$
    if it can be constructed as
    \begin{align} \label{eq:treeSB}
        G = \sum_{\bep \in B(\tau)} W_{\bep} \delta_{\theta_{\bep}}, 
    \end{align}
    where the set $B(\tau) \subset E^*$ indexes $\tau$'s leaf nodes, the random weights $\{W_{\bep}\colon \bep \in B(\tau)\}$ are constructed according to \eqref{eq:genweights}, the splitting variables $V_{\bep} \sim F_{\bep}$ and satisfy $V_{\emptyset} \perp \{V_0, V_1\} \perp \{V_{00}, V_{01}, V_{10}, V_{11}\} \perp \cdots$, and the atoms $\theta_{\bep} \ind G_0$ and are generated independently of the splitting variables $V_{\bep}$.
    Such a measure $G$ is denoted by $G \sim \prior(G_0, \{F_{\bep}\}, \tau)$.
\end{definition}
As with traditional stick-breaking, a random measure $G \sim \prior(G_0, \{F_{\bep}\}, \tau)$ turns the mixture density $\int_{\Theta} h(\cdot; \theta) \dif G(\theta)$ into the discrete sum $\sum_{\bep \in B(\tau)} W_{\bep} h(\cdot; \theta_{\bep})$.

\subsection{Tree stick-breaking with covariates}
\label{sec:methods2}

Following the well-known strategy of 
\cite{Maceachern2000}, we next extend the covariate-independent tree stick-breaking prior \eqref{eq:treeSB} by replacing the splitting variables $\{V_{\bep}\}_{\bep \in \EE^*}$ in \eqref{eq:genweights} with a sequence of stochastic processes $\{V_{\xx,\bep} \colon \xx \in \iset\}_{\bep \in \EE^*}$, where $\iset$ is a set of covariates and each splitting variable $V_{\xx,\bep}$ has distribution $F_{\xx,\bep}$.
The resulting random measure $G_{\xx} \sim \prior(G_0, \{F_{\xx,\bep}\}, \tau)$ now depends on $\xx$ through its weights:
$G_{\xx} = \sum_{\bep \in B(\tau)} W_{\xx,\bep} \delta_{\theta_{\bep}}$ where $\theta_{\bep} \ind G_0$.
As in the ``common-atoms'' or ``single-atoms'' model \citep[see e.g.,][for a review]{Quintana2022}, here we do not incorporate dependence into the atoms (nor in the tree structure).

There are a number of possible strategies to incorporate covariate dependence on weights, including utilizing probit and logit transform on the weights as is commonly done for traditional stick-breaking. In particular, we consider the logit approach due to the computational convenience for posterior inference that follows from the P\'olya-Gamma augmentation technique detailed in Section~\ref{sec:samplerfixedeffects}, though the theoretical properties we establish in Section~\ref{sec:moments} do not assume this model choice and apply more generally.
Specifically, we adopt the following logit-normal model on each splitting variable:
\begin{equation}
V_{\xx,\bep} = \text{logistic}(\eta_{\xx,\bep}), \qquad 
\eta_{\xx,\bep} = \bpsi(\xx)^\top \bgamma_{\bep},  \qquad 
\bgamma_{\bep} \sim N_R(\bmu_{\bgamma}, \bSigma_{\bgamma})
\label{eq:nu}
\end{equation}
with hyperparameters $\bmu_{\bgamma}$ and $\bSigma_{\bgamma}$, where $\text{logistic}(z) = \exp\{z / (1+z)\}$ and $\eta_{\xx,\bep}$ is a linear combination of selected functions of the covariates $\bpsi(\xx) = \{\psi_1(\xx), \ldots, \psi_R(\xx)\}^\top$.
Thus $\{\eta_{\xx,\bep}\colon \xx \in \iset\}$ is a Gaussian process with mean $\bpsi(\xx)^\top \bmu_{\bgamma}$ and covariance $\Cov(\eta_{\xx,\bep}, \eta_{\xx',\bep}) = \bpsi(\xx)^\top \bSigma_{\bgamma} \bpsi(\xx')$.
The remainder of the paper (except for Section~\ref{sec:moments}) assumes the logit-normal prior \eqref{eq:nu}.

\section{Impact of the tree on cross-covariate correlation}
\label{sec:priorcorrelation}

This section examines the impact of tree structure on the prior cross-covariate correlation between two random measures created by stick breaking with dependent mixture weights and independent atoms. 
We provide expressions for various moments of the covariate-dependent random measures and create a simulation study that explores the cross-covariate correlation between random measures.

\subsection{Moments of random measures}
\label{sec:moments}

The prior \eqref{eq:nu} satisfies the conditions Theorems \ref{thm:meanvarcorr} and \ref{thm:summoment2} place on the splitting variables.

\begin{theorem}
\label{thm:meanvarcorr}
For some $K \in \{1, 2, \ldots, \infty\}$, suppose for any covariates $\xx$ that the random measure
$G_{\xx} = \sum_{k=1}^K W_{\xx,k} \, \delta_{\theta_k}$ 
on $\Theta$ is constructed by drawing each $\theta_k$ independently from a base measure $G_0$ (that does not depend on $\xx$) on $\Theta$, and
drawing a weight vector $(W_{\xx,1}, \ldots, W_{\xx,K})$
according to some distribution (that might depend on $\xx$) on the probability simplex $\Delta_K$.

For any measurable sets $A, A' \in \mathcal{B}$ and covariates $\xx, \xx'$, we have
\begin{align}
\E(G_{\xx}(A)) &= G_0(A)  \label{eq:mean}  \\
\Var(G_{\xx}(A)) &=  \left\{G_0(A) - G_0^2(A)\right\} a_{\xx,\xx}  \label{eq:variance} \\
\Cov(G_{\xx}(A), G_{\xx}(A')) &= \left\{G_0(A \cap A') - G_0(A)G_0(A')\right\} a_{\xx,\xx}  \label{eq:covAA}  \\
\Cov(G_{\xx}(A), G_{\xx'}(A)) &= \left\{G_0(A) - G_0^2(A)\right\} a_{\xx,\xx'}  \label{eq:covxx}
\end{align}
where $a_{\xx,\xx'} =\sum_{k=1}^K \E (W_{\xx,k} W_{\xx',k})$.
If also $G_0(A) \{1 - G_0(A)\} G_0(A') \{1 - G_0(A')\} \neq 0$,
then
\begin{align}
\corr(G_{\xx}(A), G_{\xx}(A')) &= \frac{G_0(A \cap A') - G_0(A)G_0(A')}{\left[G_0(A) \{1 - G_0(A)\} G_0(A') \{1 - G_0(A')\}\right]^{1/2}} \label{eq:corrAA} \\
\corr(G_{\xx}(A), G_{\xx'}(A)) &= a_{\xx,\xx'} (a_{\xx,\xx} a_{\xx',\xx'})^{-1/2}.  \label{eq:corrxx}
\end{align}
\end{theorem}

We note that the structure of these prior moments in Theorem~\ref{thm:meanvarcorr} is identical to that of homogeneous normalized random measures \citep[cfr. Proposition 1 in][]{james2006conjugacy}.
Also, the correlations \eqref{eq:corrAA} and \eqref{eq:corrxx} discussed in this paper are always nonnegative, but negative correlations can be produced by inducing repulsion into the atom-generating measure for exchangeable observations \citep{petralia2012repulsive} or partially exchangeable observations \citep{ascolani2023nonparametric}.

Per Theorem~\ref{thm:meanvarcorr}, each of these various moments factorizes into a function of the base measure $G_0$ and a function of the quantity $a_{\xx,\xx'}$.
The mean \eqref{eq:mean} implies $G_0$ can be viewed as the mean of the random measure $G_{\xx}$ while the correlation \eqref{eq:corrAA} does not depend on tree depth.
On the other hand (and of greater interest), the cross-covariate correlation \eqref{eq:corrxx} depends not on the base measure $G_0$ but rather on $a_{\xx,\xx'}$, which Theorem~\ref{thm:summoment2} expresses as a function of $\tau$ and the splitting variables' mean and cross-covariate covariance.
In particular, this theorem states that $a_{\xx,\xx'}$ for a balanced tree approaches zero as $K \rightarrow \infty$ whereas $a_{\xx,\xx'}$ for a lopsided tree 
\citep[which is mostly derived in Appendix~2 of ][]{Rodriguez2011} 
approaches a positive limit, which means a lopsided tree induces a baseline cross-covariate \textit{covariance} value \eqref{eq:covxx} between random measures that does not vanish as the number of weights approaches infinity.
Corollary~\ref{corollary:lowerbounds} states these statements also apply to their cross-covariate \textit{correlation} counterparts \eqref{eq:corrxx} if the splitting variables have mean $1/2$ and nonnegative cross-covariate correlation.

\begin{theorem}
\label{thm:summoment2}
Suppose a set of weights $\{W_{\xx,\bep} \colon \bep \in B(\tau)\}$ is constructed by stick-breaking according to tree structure $\tau$, where $B(\tau)$ is the set of leaf nodes in $\tau$ and $K = |B(\tau)|$ is the number of leaf nodes. 
Also assume that the distribution of any splitting variable $V_{\xx,\bep}$ does not depend on $\bep$.
Let $a_{\xx,\xx'} = \sum_{\bep \in B(\tau)} \E (W_{\xx,\bep} W_{\xx',\bep})$. 
If $\tau$ is a lopsided tree, then, letting $e_{\xx,\xx'}=\E (V_{\xx}) + \E (V_{\xx'}) - \E (V_{\xx}V_{\xx'})$, we have
\begin{align*}
a_{\xx,\xx'} 
&= \frac{\E (V_{\xx} V_{\xx'})}{e_{\x,\x'}} + \bigg(1-\frac{\E (V_{\xx} V_{\xx'})}{e_{\x,\x'}}\bigg) (1-e_{\x,\x'})^{K-1} \xrightarrow[K \rightarrow \infty]{} \frac{\E (V_{\xx} V_{\xx'})}{e_{\x,\x'}}. 
\end{align*}
If instead $\tau$ is a balanced tree and $m = \log_2 K$ is a nonnegative integer, then
\begin{equation*}
a_{\xx,\xx'} = \left\{1 - \E (V_{\xx}) - \E (V_{\xx'}) + 2 \E (V_{\xx} V_{\xx'})\right\}^m \xrightarrow[m \rightarrow \infty]{} 0.
\end{equation*}
\end{theorem}

The two values of $a_{\x,\x'}$ above agree for $K \in \{1,2\}$, which reflects the equivalence of any binary tree for these two values of $K$.

\begin{corollary}[Bounds for \eqref{eq:corrxx}]
\label{corollary:lowerbounds}
Given the assumptions in Theorem~\ref{thm:summoment2}, suppose also that $\E (V_{\xx}) = 1/2$ for any $\xx$ and $\Cov(V_{\xx}, V_{\xx'}) \geq 0$ for any $\xx,\xx' \in \iset$. 
\begin{enumerate}
    \item[(a)] Lower bounds for \eqref{eq:corrxx} are $1/3 + (2/3)4^{1-K}$ for a lopsided tree and $2^{-m}$ for a balanced tree.
    \item[(b)] If for some $(\xx,\xx')$ the conditions $\Var(V_{\xx}) \Var(V_{\xx}')>0$ and $\corr(V_{\xx}, V_{\xx'}) < 1$ are also satisfied, then the cross-covariate correlation \eqref{eq:corrxx} for a balanced tree and this $(\xx,\xx')$ shrinks to zero as $m \rightarrow \infty$. 
    \item[(c)] For such $(\xx,\xx')$ as in part (b), the cross-covariate correlation \eqref{eq:corrxx} for the lopsided tree is strictly larger than that for the balanced tree 
    with the same number of leaves when the number of leaves is sufficiently large.
\end{enumerate}
\end{corollary}

The statement in part (c) applies to sufficiently large trees and so in principle the lopsided tree could lead to weaker cross-covariate correlation \eqref{eq:corrxx} than that for the balanced tree, but we found empirically that this would require the correlation $\corr(V_{\xx}, V_{\xx'})$ to be extremely close to~1 and the trees to have very few leaves (e.g., $K=4$ or 8), and even then the lopsided tree would have only slightly smaller cross-covariate correlation.
(See the Supplemental Material for a numerical experiment that validates this claim.)

We will further explain and discuss the above corollary in Sections \ref{sec:corrGxGy} and \ref{sec:choiceofnumberofleaves}.

Finally, the following theorem establishes the important property that the random measure $G_{\xx}$ changes smoothly with respect to $\xx$.
This property
requires a moment condition which is satisfied if the stochastic process $\{V_{\xx} \colon \xx \in \iset\}$ is second-order stationary.

\begin{theorem}[Smoothness]
\label{thm:smooth}
Given the assumptions in Theorem~\ref{thm:summoment2}, if both $\E (V_{\xx'}) \rightarrow \E (V_{\xx})$ and $\E (V_{\xx}V_{\xx'}) \rightarrow \E V_{\xx}^2$ as $\xx' \rightarrow \xx$, then \eqref{eq:corrxx} $\rightarrow 1$ as $\xx' \rightarrow \xx$.
\end{theorem}

\subsection{Numerical illustration on prior cross-covariate correlation}
\label{sec:corrGxGy}

Here we explore the impact of the cross-covariate correlation between splitting variables on the cross-covariate correlation between random measures:
given covariates $\xx$ and $\xx'$, how does $\corr(V_{\xx}, V_{\xx'})$ affect $\corr\left(G_{\xx}(A), G_{\xx'}(A)\right)$ for any measurable set $A \in \mathcal{B}$?
We provide insight into this question through the following example.

\begin{example}
We reduce the number of influences on $\corr(V_{\xx}, V_{\xx'})$ by making the following assumptions:
$\bmu_{\bgamma} = 0_2$, 
$\bSigma_{\bgamma} = \text{diag}(\sigma^2_1, \sigma^2_2)$,
$\sigma^2_1 > 0$, 
$\bpsi(\xx) = (1,0)^\top$, and
$\bpsi(\xx') = (1,1)^\top$.
Though seemingly strict, these assumptions encompass a large class of scenarios.
The mean-zero assumption is reasonable if no prior information is given.
These assumptions also imply $\corr(V_{\xx}, V_{\xx'}) = \corr(\eta_{\xx}, \eta_{\xx'}) = (1 + \tilde{\sigma}^2_2)^{-1/2}$,
which is a strictly decreasing function of $\tilde{\sigma}^2_2 \coloneqq \sigma^2_2/\sigma^2_1 \geq 0$ whose image is $(0,1]$.
Thus, any positive value of $\corr(V_{\xx}, V_{\xx'})$ can be achieved by using the appropriate $\tilde{\sigma}^2_2$ value if $\xx \neq \xx'$.
Similarly, these assumptions reduce $\corr(G_{\xx}(A), G_{\xx'}(A))$ (whose expression is provided by \eqref{eq:corrxx} and Theorem~\ref{thm:summoment2}) to a function of $\sigma_1^2$, $\sigma_2^2$, $K$, and $\tau$.

\begin{figure}
\begin{center}
\includegraphics[width=\textwidth]{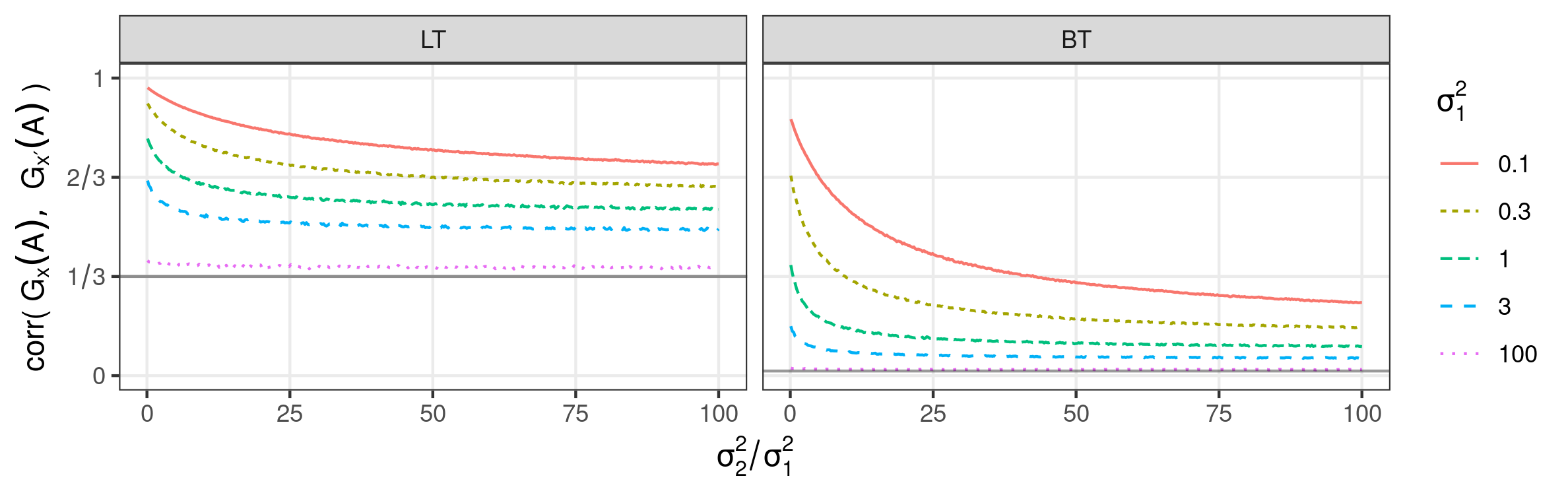}
\end{center}
\caption{Simulated values of \eqref{eq:corrxx} as a function of $\tilde{\sigma}^2_2 \coloneqq \sigma^2_2/\sigma^2_1$ for a lopsided and balanced tree each with $K=64$ and various values of $\sigma^2_1$. Lower bounds are shown as solid horizontal lines.}
\label{fig:GxGy}
\end{figure}

Figure~\ref{fig:GxGy} shows the behavior of $\corr(G_{\xx}(A), G_{\xx'}(A))$, which by Corollary~\ref{corollary:lowerbounds} has lower bounds of $1/3$ and $1/K$ for, respectively, a lopsided tree and balanced tree.
These lower bounds hold regardless of the degree of correlation between splitting variables, which in this scenario is controlled by choice of $\tilde{\sigma}^2_2$.
Thus, the lopsided-tree scheme always imposes a nontrivial baseline correlation between random measures while the balanced-tree scheme can achieve both large and small correlation values, 
which provides more flexibility in setting prior correlation values.
\end{example}

\subsection{Choice of number of leaves}
\label{sec:choiceofnumberofleaves}

This section considers how a practitioner should choose the number of leaves $K$ in the tree $\tau$ used in a tree stick-breaking prior. 
First, we consider the scenario where the splitting variables follow the conditions in Theorem~\ref{thm:summoment2}.
As done in \cite{argiento2022infinity}, we emphasize the distinction between $K$ and the number of inferred clusters.
The chosen $K$ should be large enough to capture the ``true'' number of clusters, but not so large that the posterior inference algorithm in Section~\ref{sec:samplerfixedeffects} becomes computationally intractable.
We believe that $K$ in the range between 16 and 64 should serve as recommended default values for most applications, which strikes a balance between flexibility and computational efficiency. For the example in Section~\ref{sec:corrGxGy} $K=64$ allows the balanced-tree to achieve any prior cross-covariate correlation value \eqref{eq:corrxx} above $1/K \approx 0.016$.
Another option would be to set $K$ to be the smallest power of two greater than or equal to twice the prior expected number of clusters; this prior information can be obtained either through domain knowledge or through applying e.g. $k$-means clustering \citep{hartigan1979algorithm}.
Regarding robustness of inference with respect to $K$, Theorem~\ref{thm:summoment2} suggests that \eqref{eq:corrxx} can be sensitive to choice of $K$ for either tree structure if $K$ is small. At the other extreme, for a lopsided tree the exponential bounds in the finite approximation theorems of \cite{Ishwaran2002approximate} imply the value of $K$ beyond say $64$ seems to affect inference by a trivial amount (assuming $K$ upper bounds the ``true'' number of clusters). 
We introduce Theorem~\ref{thm:BTagree} to make a similar statement for a balanced tree.

\begin{theorem} \label{thm:BTagree}
Consider a sequence $\{ \T_j = \{A_{\bep}: \bep \in \{0,1\}^j\}: j \in 0,1,2,\ldots\}$ 
of measurable partitions of the sample space obtained by splitting every set in the preceding partition into two new sets.
Suppose the sample space is a metric space and $\text{diam}(A_{\bep}) = 2^{-|\bep|}$ for all $A_{\bep}$, where $|\bep|$ is the length of the string $\bep$ of 0s and 1s.
If two tail-free processes agree on all subsets $A_{\bep}$ with $|\bep| \leq M$ for some positive integer $M$, 
then the Wasserstein distance between these two processes is bounded above by $2^{-M+1}$. 
\end{theorem}

Putting aside computational implications, it is natural to consider the theoretical feasibility of constructing a suitable stick-breaking model on a balanced-tree of infinite depth. We believe this is possible if the splitting variables are specified with extra care. To see this, note that had one simply adopted i.i.d.\ splitting variables as we have done so far, then the resulting model would have zero prior probability to generate any non-trivial cluster sizes and hence lead to the nonsensical inference that there are {\em always} as many distinct clusters as there are distinct observations in the data. This is in stark contrast to lopsided-tree stick-breaking models such as those that give rise to the Dirichlet process for which i.i.d.\ splitting variables suffice.
Strategies to regularize infinitely-deep balanced trees through prior specification have been well studied in the density estimation context for tail-free processes \citep{freedman1963asymptotic,freedman1965asymptotic,fabius1964asymptotic} and in particular for Pólya trees \citep{lavine1992some,lavine1994more}. In those applications, the prior variance of the splitting variables must decrease sufficiently fast down the tree. Interestingly the consideration for specifying the splitting variable here is just the {\em opposite}. Prior specifications that produce well-defined densities will in fact again fall into the trap of generating no non-trivial clusters with probability~1. In fact, for the infinite balanced-tree stick-breaking process to work as a functional prior for cluster weights, the splits must increasingly resemble Bernoulli distributions sufficiently fast down the tree
so that most stick pieces essentially remain unchanged in deep enough levels of the tree. 
We defer the study of the specific conditions for specifying the splitting variables to future work. 

An alternative strategy for an infinite balanced tree is to treat the depth of the tree, or more generally the specific topology of the tree, as an unknown quantity of interest and place a prior on it, resulting in the counterpart of a mixture of finite mixtures \citep{antoniak1974mixtures,richardson1997bayesian,nobile2004posterior,gnedin2006exchangeable,gnedin2010species,de2013asymptotic,Miller2018,grazian2020loss,fruhwirth2021generalized}.  
In this case, inferring the number of clusters along with the allocation of the clusters over the leaves becomes in essence a problem of learning an unknown tree topology. Similar strategies have also been employed in the density estimation context based on P\'olya trees and related models \citep{Wong2010,Ma2017}, but the computational task of learning the unknown tree is less demanding for those tree-based density models because not only can the conditional model there {\em given} the tree topology often be analytically integrated out but the Bayes factor between two slightly different tree topologies can often be computed analytically involving only a small subset of the training data. Neither is true in the context of modeling cluster sizes. Computational techniques developed for tree learning and also regression in density models, though immediately available in theory, are not practically feasible and what type of computational strategies are effective remains an open and interesting question. We defer further investigation in this direction also to future work. The remainder of this paper will focus on the simple case of a finite fixed $K$ along with splitting variables whose distribution depends on the covariates in a node-specific fashion but does not depend on the tree node it belongs to in any other way.

\section{Impact of the tree on posterior uncertainty}
\label{sec:posterioruncertainty}

This section begins by detailing the posterior computation of the mixture model $f(\cdot; G_{\xx}) = \int_{\Theta} h(\cdot; \theta) \dif G_{\xx}(\theta)$
with a tree stick-breaking prior on the mixing measure $G_{\xx}$.
If $\tau$ is a lopsided tree with $K < \infty$ leaves, we call the resulting mixture a finite lopsided-tree mixture.
We similarly define a finite balanced-tree mixture, where $K$ must be a power of $2$.
This section then presents a case study that analyzes flow-cytometry data and illustrates the impact of tree structure on the posterior uncertainty of covariate effects in mixture weights. 

\label{sec:samplerfixedeffects}

For posterior computation, we generalize \cite{Rigon2021}'s Gibbs sampler to admit any stick-breaking scheme. 
Their Gibbs step for the regression coefficients $\bgamma$ relies on \cite{Polson2013}'s P\'olya-Gamma data augmentation technique, which allows efficient posterior sampling of a Bayesian logistic regression.
For this technique, \cite{Polson2013} carefully construct the P\'olya-Gamma family of distributions to allow conditionally conjugate updating for the coefficient parameter and provide a fast, exact way to simulate P\'olya-Gamma random variables.
They show a posterior sampler for $\bgamma$ is obtained by iterating between a step that, conditional on $\bgamma$, samples the P\'olya-Gamma data and a step that, conditional on the P\'olya-Gamma data and regression responses, samples $\bgamma$ from a multivariate Gaussian distribution.

Our generalization of this Gibbs step is stated explicitly in Algorithm~1 in the Supplemental Material (we later introduce Algorithm~\ref{alg:gibbsrandomeffect} in the main text which is a further generalization that incorporates both fixed and random effects). 
Given a posterior draw, for all $i = 1, \ldots, n$ let $C_{\tau}(i)$ be $\tau$'s leaf node assigned to observation $i$.
For each internal node $\bep$, the update for the coefficient $\bgamma_{\bep}$ relies on a Bayesian logistic regression
with responses $Z_{i\bep}$, defined as the indicator that leaf node $C_{\tau}(i)$ is a ``left descendant'' of node $\bep$, 
for all $i$ corresponding to a descendant node of $\bep$.
A ``left descendant'' of $\bep$ is either $\bep$'s left child or a descendant of $\bep$'s left child, and ``right descendant'' is defined similarly.
In a lopsided tree the left child of any internal node is a leaf (see Figure~\ref{fig:treestructure}) whereas for any internal node in a balanced tree the number of left descendants equals the number of right descendants.
This formulation easily applies to any finite tree stick-breaking scheme.
However, we find that training a balanced-tree mixture model takes less time than training its lopsided-tree counterpart for the data sets in Section~\ref{sec:grifols} and Section~3 in the Supplementary Material; this observation is supported by the theoretical discussion in Section~2 in the Supplementary Material.

For many applications, it is crucial to include random effects into the splitting-variable model. 
Without random effects, the model \eqref{eq:nu} makes the strong assumption that mixture-weight differences between groups is due entirely to differences in covariates. 
A mixture model that assumes \eqref{eq:nu} will resolve any large difference in cluster proportions between the two individuals that share the same covariates by breaking up the would-be cluster into many smaller clusters; such a mixture model would thus infer many more clusters than actually exists in the data (as we have seen from experience). 
Section~\ref{sec:modelfitting} provides details of the specific random effects and the Gibbs sampler for a flow-cytometry case study, but these can easily generalize to other contexts.

\subsection{The impact of the tree on posterior inference: a case study}
\label{sec:grifols}

We conduct a case study involving covariate-dependent clustering to demonstrate the influence of the tree structure on the posterior inference of the covariate effects over the cluster sizes. The scientific objective in this case study is to quantify the impact of an African-American female's age on her proportions of T-cell types. Our analysis uses two groups of individuals, younger (aged 18-29) and older (aged 50-65), from a publicly available data set to establish normative ranges \citep{Yi2019} using the Human Immunology Profiling Consortium T cell immunophenotyping panel \citep{RN2}. This panel has antibodies to cell surface proteins, known as biomarkers, designed to identify CD4+ and CD8+ T cell activation and maturational status but is not specialized to resolve other immune cell types or degrees of immune senescence. 
The sample is of all peripheral blood mononuclear cells, and clusters may include non-T cell subsets.
In standard analysis, an expert is required to visually identify distinct cell subsets using a sequence of 2D boundaries known as gates. 
We will instead identify cell subsets using a mixture model. 

Our analysis is based on flow cytometry data measured on blood samples from 15 healthy plasma donors, six of which are 18-29 years old and remaining nine of which are 50-65 years old. Each sample can be roughly considered a collection of exchangeable observations (each corresponding to a blood cell) from a seven-dimensional continuous sample space, with each dimension corresponding to the measurement from one marker. These 15 subjects together produce too many viable cells for the model to fit in a reasonable amount of time with Markov chain Monte Carlo. As such, we subset the data in a way that uses all 15 subjects while representing each age group by the same number of viable cells. 
Hence we subset a total of $n=403200$ viable cells where each age group is represented by $n/2$ cells.
Within each age group, each subject contributes the same number of cells, i.e. the six 18-29 subjects each contributes one-sixth of the $n/2$ cells and the nine 50-65 subjects each contributes one-ninth of the $n/2$ cells.

\subsection{A mixed-effects model and a recipe for Bayesian computation}
\label{sec:modelfitting}

In this study, cells are grouped by the subject they come from and subjects are grouped by the laboratory in which their cells are collected.
We account for any resulting group effects in the mixture weights by including random effects in the splitting variables.
For each subject $s$, each splitting variable will include fixed effects $\bfe$ for covariates $\bpsi(\xx_s) = (1, \text{age group of subject }s)^\top$, a random effect $\res_s$ for the subject, and a random effect $\rej_j$ for the subject's batch $j$ (we omit $j$'s dependence on $s$ to avoid visual clutter): 
\begin{align}
\text{logit } V_{s,\ep} &= \bpsi(\xx_s)^\top \bfe_{\ep} + \rej_{j,\ep} + \res_{s,\ep}.
\label{eq:mixedeffectsmodel}
\end{align}
The covariates and fixed effects are treated as in \eqref{eq:nu}, and the random effects have priors
\begin{align*}
\rej_{j,\ep} | \rep_{\ep}^{(\rej)} \ind \Normal(0, (\rep_{\ep}^{(\rej)})^{-1}), \quad
\res_{s,\ep} | \rep_{\ep}^{(\res)} \ind \Normal(0, (\rep_{\ep}^{(\res)})^{-1}), \quad 
\rep_{\ep}^{(\rej)}, \rep_{\ep}^{(\res)} \ind \text{Gamma}(1, 1).
\end{align*}
Using \cite{Wang2018analysis}'s two-block Gibbs sampler, Algorithm~\ref{alg:gibbsrandomeffect} extends our Gibbs step from Algorithm~1 in the Supplementary Material to also update the random effects.

\begin{algorithm}
\SetAlgoLined
\KwResult{Update each fixed-effects $R$-tuple $\bgamma_{\bep}$ and random-effects $(J+S)$-tuple $\uu_{\bep}$.}
\For{each internal node $\bep$ in binary tree $\tau$}{ 
    Let $\cD_{\bep} \subseteq \{1,\ldots,n\}$ be the set of indices $i$ where $C_{\tau}(i)$ is a descendant of node $\bep$ \;
    Update precision parameter $[\rep_{\bep}^{(\rej)} \mid \cdots] \sim \Gamma (a_{\bep}^{(\rej)} + 0.5, b_{\bep}^{(\rej)} + 0.5 \sum_{j \in \cJ_{\bep}} \rej_{j,\bep}^2)$, where $\cJ_{\bep} \subseteq \cJ$ is the set of batches corresponding to observations in $\cD_{\bep}$ \;
    Update precision parameter $[\rep_{\bep}^{(\res)} \mid \cdots] \sim \Gamma (a_{\bep}^{(\res)} + 0.5, b_{\bep}^{(\res)} + 0.5 \sum_{s \in \cS_{\bep}} \res_{s,\bep}^2)$, where $\cS_{\bep} \subseteq \cS$ is the set of subjects corresponding to observations in $\cD_{\bep}$ \;
    \For{every observation $i \in \cD_{\bep}$} {
        Sample $[\omega_{i\bep} \mid \cdots] \sim \text{P\'olya-Gamma}(1, \, |\bpsi(\xx_s)^\top \bfe_{\ep} + \res_{s,\ep} + \rej_{j,\ep}|)$, for the subject $s$ and batch $j$ associated with $i$ \;
    }
    Let $\cL_{\bep} \subset \cD_{\bep}$ be the set of indices corresponding to left descendants of node $\bep$ \;
    Update fixed and random effects by drawing from the full conditional 
    \begin{align*}
        \left[ \begin{pmatrix} \bgamma_{\bep} \\ \uu_{\bep} \end{pmatrix} \mid \cdots \right] \sim \text{N}_{R+J+S}(\bmu_{\bgamma_{\bep}, \uu_{\bep}}, \bSigma_{\bgamma_{\bep}, \uu_{\bep}})
    \end{align*}
    where, letting $\bM_{\bep}$ be the $|\cD_{\bep}| \times (R+J+S)$ matrix with row entries $(\bpsi(\xx_i)^\top, \z_i^\top)$ for only those $i \in \cD_{\bep}$, $\bkappa_{\bep} = (1_{i \in \cL_{\bep}} - 0.5)_{i \in \cD_{\bep}}$, $\bOmega_{\bep} = \text{diag}(\omega_{i{\bep}}: i \in \cD_{\bep})$, and $\bRep_{\bep} = (\rep_{\bep}^{(\rej)} \I_J) \oplus (\rep_{\bep}^{(\res)} \I_S)$, 
    \begin{align*}
        \bmu_{\bgamma_{\bep}, \uu_{\bep}} &= \bSigma_{\bgamma_{\bep}, \uu_{\bep}} \left[\bM_{\bep}^\top \bkappa_{\bep} + \begin{pmatrix} \bSigma_{\bgamma}^{-1} \bmu_{\bgamma} \\ 0_{(S+J) \times 1} \end{pmatrix} \right], \\
        \bSigma_{\bgamma_{\bep}, \uu_{\bep}}^{-1} &= \bM_{\bep}^\top \bOmega_{\bep} \bM_{\bep} + (\bSigma_{\bgamma}^{-1} \oplus \bRep_{\bep}).
    \end{align*}
}
\caption{Gibbs step to update fixed and random effects under any binary tree.}
\label{alg:gibbsrandomeffect}
\end{algorithm}

To the data we fit a lopsided-tree mixture model and a balanced-tree mixture model each with skew-normal kernels and hyperparameter values of $K=16$, prior mean $\bmu_{\gamma} = 0_R$, and prior covariance $\bSigma_{\gamma} = 10 I_R$ where $R=2$.
Each chain burns in $5000$ steps before sampling every $10$ steps to ultimately keep $1000$ posterior draws.
Both models use \textit{cross-sample calibration} to account for subject-data being collected in different batches \citep{gorsky2023coarsened}.

\subsection{Posterior summaries of the covariate effects on mixing weights}

Because our scientific interest concerns the \textit{general} African-American female population rather than only the subjects in our study, population-level parameters such as the effects of age should retain nontrivial posterior uncertainties due to the relatively small number of subjects in the study. 
Hence the effects of the study's subjects on the mixture weights should be captured by the mixture model but the covariate effects on the cluster sizes should only be quantified by the difference in mixture weights across covariates at the {\em population} level, not for the specific individuals in the study.
We compute this {\em population-level} weight difference by first fitting a mixture model with the mixed-effects model \eqref{eq:mixedeffectsmodel}. 
Using this fitted mixture model, for each subject $s$ and internal node $\bep$ we then compute a {\em population-level} splitting variable $V_{s,\ep}^*$ using the posterior draws of only the covariate fixed effects (i.e., we omit the posterior draws of the random effects $u_s$ and $v_j$ in this computation):
\begin{align*}
\text{logit } (V_{s,\ep}^*)^{(d)} &= \bpsi(x_s)^\top \bfe_{\ep}^{(d)} + 0\rej_{j,\ep}^{(d)} + 0\res_{s,\ep}^{(d)}, \quad d=1,\ldots,1000,
\end{align*}
where $d$ indexes the $1000$ posterior draws;
this population-level splitting variable hence depends on the subject $s$ only through its age group. 
For each age group $x$, we convert $x$'s splitting variables into {\em population-level} mixture weights $\bfW_x^* \coloneqq (W_{x,\bep_1}^*, \ldots, W_{x,\bep_K}^*) \in \Delta_K$ using the usual tree-dependent conversion \eqref{eq:genweights}.
We can then measure the age effect on the mixture weights at the population level as 
\begin{align} \label{eq:grifolswtdiff}
    \bfW_{older}^* - \bfW_{younger}^*.
\end{align} 

\subsection{The influence of the tree structure on posterior inference}
\label{sec:influence}

\begin{figure}[p]
    \includegraphics[width=0.48\textwidth,valign=t]{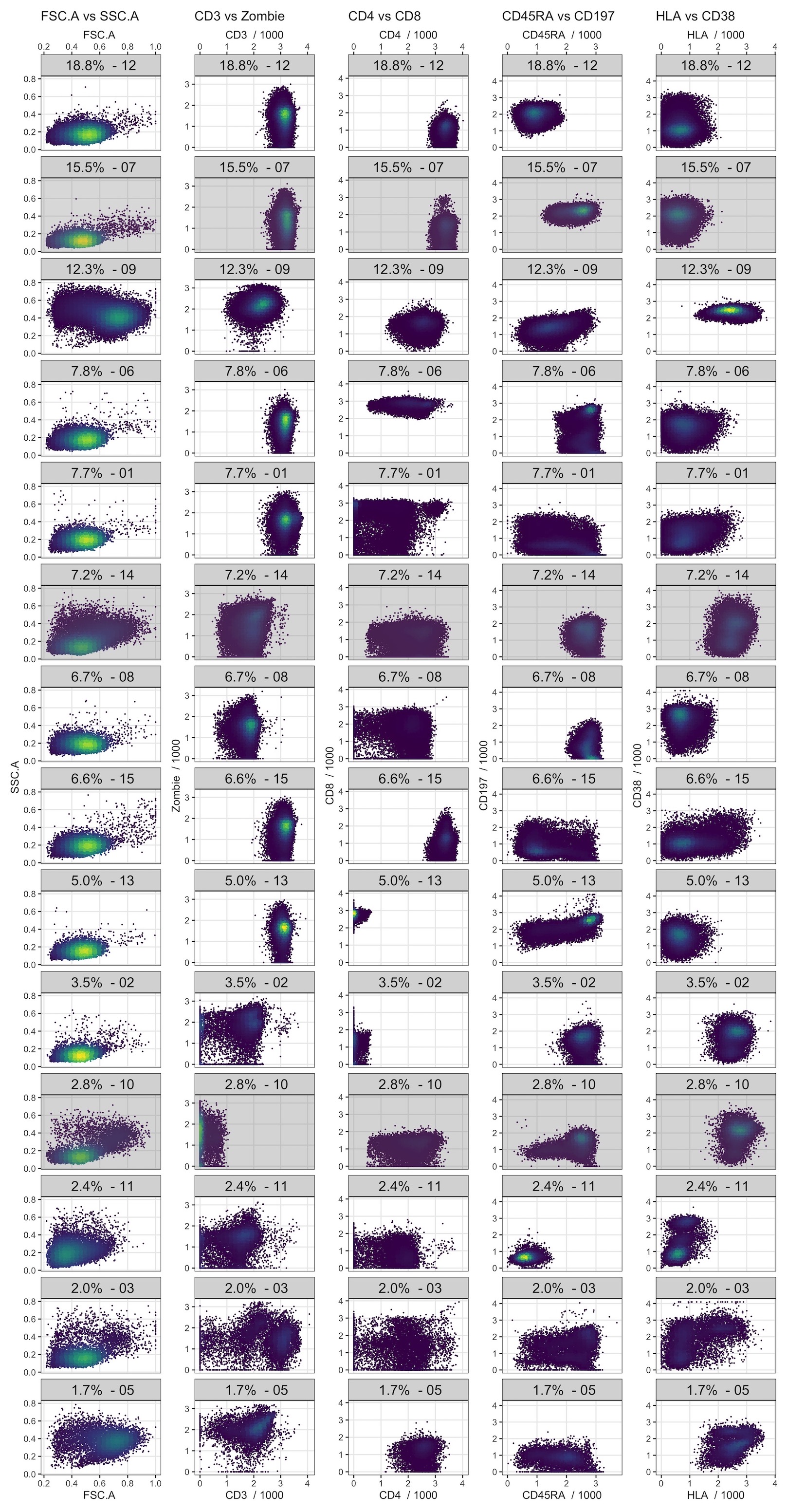}
    ~
    \includegraphics[width=0.48\textwidth,valign=t]{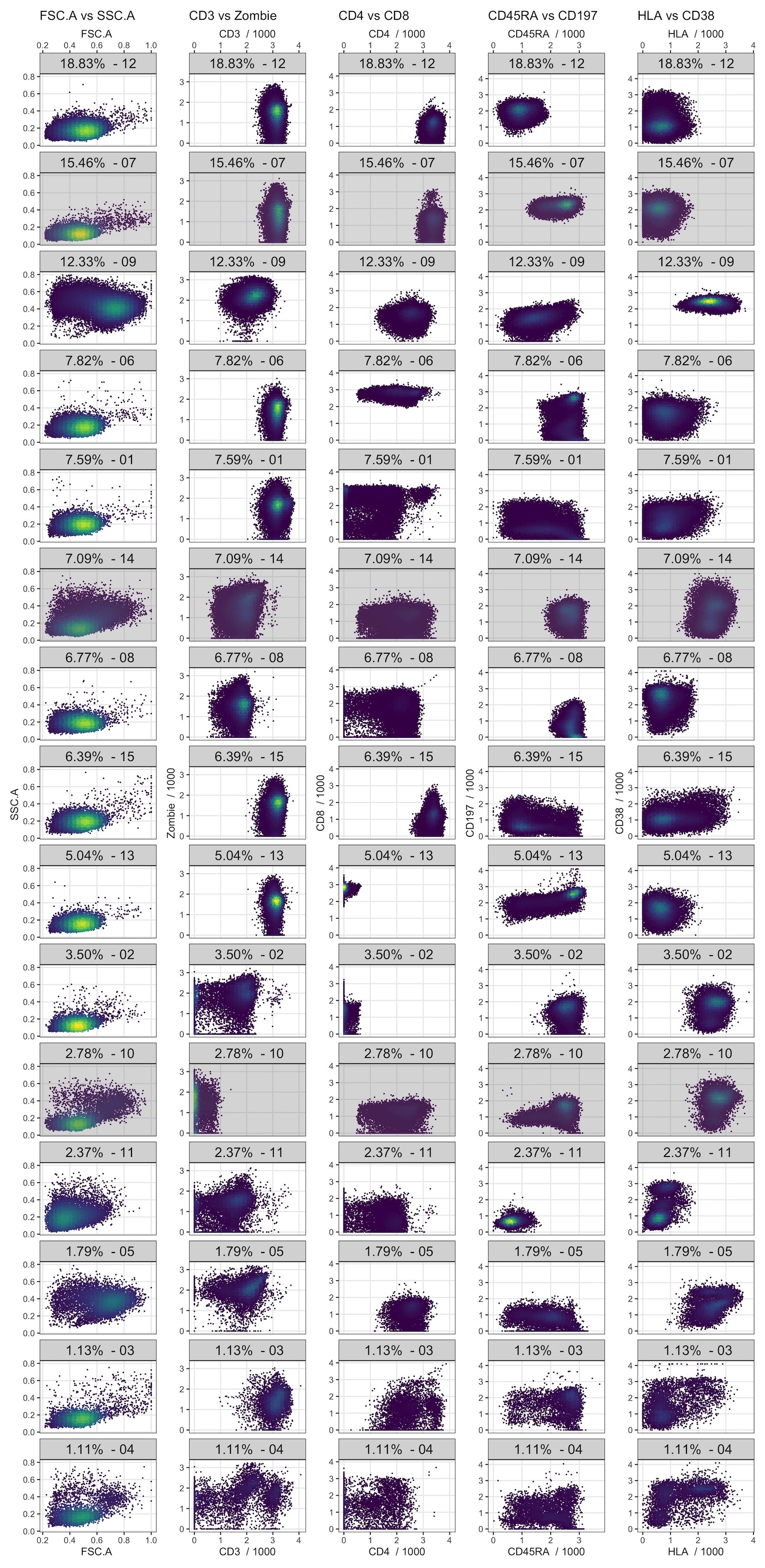}
\caption{Clusters inferred by the lopsided-tree mixture model (left) and the balanced-tree mixture model (right). The highlighted panels indicate the clusters discussed in the text.}
\label{fig:GRIFOLSclusters}
\end{figure}

After fitting the mixed-effects mixture model under both a lopsided tree and a balanced tree, we compare the inference between the corresponding posterior distributions. The two models infer very similar {\em sample-level} cluster sizes and shapes, see Figure~\ref{fig:GRIFOLSclusters}, which indicates a degree of robustness in the inference and creates an approximate one-to-one correspondence between most of the lopsided-tree clusters and most of the balanced-tree clusters. This robustness in inferring {\em sample-level} cluster sizes and shapes, i.e., those for the specific samples collected in the study, are expected for flow cytometry given the massive number of cells in each sample. The difference in the lopsided-tree mixture and balanced-tree mixture is expected to be apparent on the {\em population-level} cluster sizes given the limited number of samples and the substantial sample-to-sample variability typically observed in flow cytometry.

\begin{figure}
    \includegraphics[width=\textwidth]{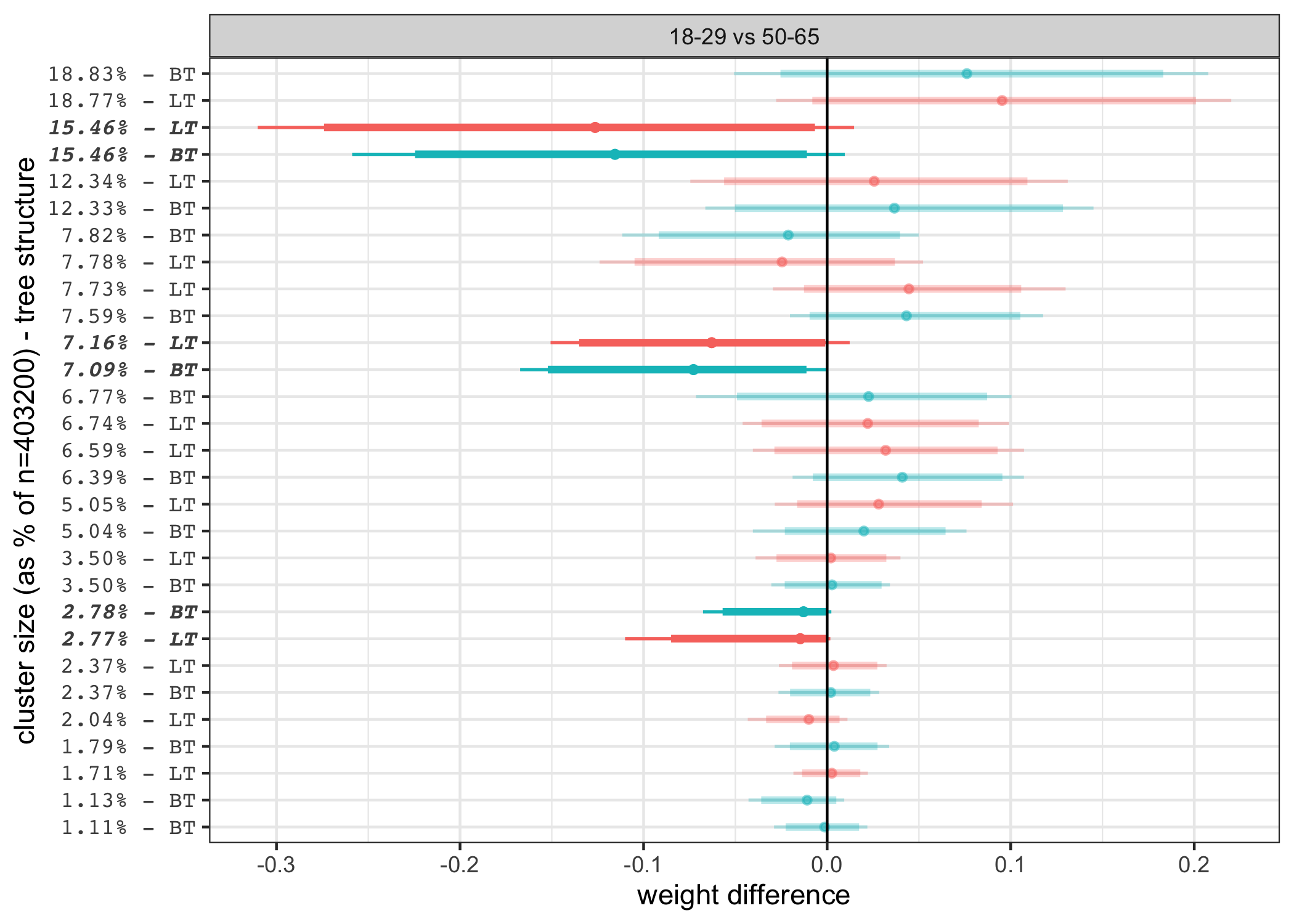}
    \caption{Credible intervals of \eqref{eq:grifolswtdiff} for both the fitted lopsided-tree mixture and balanced-tree mixture model. Thicker bar indicates $90\%$ quantile and thinner bar indicates $95\%$ quantile. The bolded italic labels indicate the clusters discussed in the text.}
    \label{fig:covdiff_withoutre_both}
\end{figure}

Interesting differences show up in the posterior distributions of population-level parameters such as the covariate effects on cluster sizes. Given the approximate correspondence in the sample-level clusters, we can directly compare the inference on covariate effects (which are on the {\em population-level} cluster sizes) between the two models.
Figure~\ref{fig:covdiff_withoutre_both} shows both sets of credible intervals of \eqref{eq:grifolswtdiff} on each cluster  in order of cluster size, which allows easier visual comparison between corresponding lopsided-tree and balanced-tree clusters. 
Even for cluster pairs with very similar size and shape, some of the corresponding credible intervals are noticeably different.
In particular, consider the three cluster pairs whose $90\%$ credible intervals are most away from zero.
For one of these three cluster pairs (whose size is roughly $7.1\%$ and whose biomarker values --- i.e., the plotted values in Figure~\ref{fig:GRIFOLSclusters} --- align with activated monocytes), the two credible intervals are similar in length but differ in location. 
For the remaining two cluster pairs (whose sizes are roughly $15.5\%$ and $2.8\%$ and whose biomarker values respectively align with those of CD4+ naïve T cells and resting monocytes), the balanced-tree credible intervals have smaller skewness and spread than do the lopsided-tree credible intervals. 
Regarding CD4+ naïve T cells, CD4+ T cells generally coordinate the overall immune response by the secretion of signaling molecules, and naïve T cells are cells which have never previously encountered antigen but might become memory cells after encountering antigen.
Hence it is highly plausible that naïve T cells would decrease with age as their production slows down markedly after adolescence \citep{RN3}, as suggested by the two credible intervals of this cluster. 

Why these credible intervals are so different is difficult to pinpoint exactly due to the many moving parts in a mixture model, but we can offer a few conjectures.
For example, consider the prior assumption of mean-zero covariate effects in the splitting variables.
For either tree, it pushes clusters with strong covariate effects toward leaf nodes graphically near each other because such a configuration would allow more splitting variables to maintain (near) zero-valued covariate effects. 
For this same reason, it also pushes such clusters away from the tree's root.
But for the lopsided-tree mixture model, prior stochastic ordering in the mixture weights pushes larger clusters \textit{toward} the root and hence competes with the previous mechanism over large clusters with strong covariate effects (such as the cluster whose size is roughly $15.5\%$).
In addition, for a mixture model with $K$ components, the lopsided-tree weights are on average a product of $\approx K/2$ splitting variables whereas each balanced-tree weight is a product of $\log_2 K$ splitting variables, which implies the lopsided-tree weights are a more interdependent function of the splitting variables than are the balanced-tree weights.
The exact way these mechanisms affect the quality of the posterior inference is unclear, but the seemingly fewer moving parts and more efficient model representation of the mixture weights in the balanced-tree mixture model are appealing.

In addition to offering the above conjectures, we also conduct the following experiment to gauge the trustworthiness of the lopsided-tree and balanced-tree inferences in the above GRIFOLS data analysis. 
Rather than use completely synthetic data, which would omit much of the complexity inherent in flow cytometry data, we instead modify the GRIFOLS data by injecting a covariate effect of known size into one of the clusters in Figure~\ref{fig:GRIFOLSclusters}. 
Because we perform many runs of this experiment, for time and memory considerations we down sample from 403200 cells to 100800 cells and have each chain burn in 5000 steps before sampling every 20 steps to ultimately keep 500 posterior draws.
However, this downsampling results in some clusters not being inferred (e.g., they might be absorbed by one or many other clusters) and hence not every cluster in Figure~\ref{fig:GRIFOLSclusters} corresponds to a cluster in the downsampled regime, and vice versa.
Furthermore, we would like the to-be-enhanced cluster to not consist of too many cells so that the induced perturbation to the data is as local as possible. 
Hence, we choose the cluster labelled $2.8\%/2.78\%$ in Figure~\ref{fig:GRIFOLSclusters}, which, after downsampling, is still clearly identified and separated from other clusters.

\begin{table}
\begin{tabular}{|r|cc|cc|cc|}
  \hline
   & \multicolumn{2}{c|}{\textbf{length median (sd)}} &\multicolumn{2}{c|}{\textbf{p truth in}} &\multicolumn{2}{c|}{\textbf{p subset}} \\
CI level & LT & BT & LT & BT & pLTinBT & pBTinLT \\ 
  \hline
0.50 & 0.036 (0.017) & 0.033 (0.018) & 0.595 & 0.550 & 0.080 & 0.165 \\ 
0.80 & 0.073 (0.032) & 0.066 (0.034) & 0.740 & 0.790 & 0.100 & 0.235 \\ 
0.90 & 0.098 (0.041) & 0.086 (0.044) & 0.910 & 0.925 & 0.120 & 0.325 \\ 
0.95 & 0.125 (0.050) & 0.106 (0.053) & 0.940 & 0.935 & 0.130 & 0.380 \\ \hline
\end{tabular}
\caption{Summary of the credible intervals for the cluster labeled $2.8\%$/$2.78\%$ in Figure~\ref{fig:GRIFOLSclusters} in the GRIFOLS simulation study detailed in Section~\ref{sec:influence}.}
\label{tbl:GRIFOLSsim}
\end{table}

Given this cluster, we aim to assess the quality of the inference to be performed.
Although we know how many of the cluster's cells come from younger and older subjects, the ``correct'' value of the mixture-weight difference \eqref{eq:grifolswtdiff} is still unclear due to the various complex properties of the data and imposed model, such as misshapen clusters and subject random effects.
Hence our data modification consists of enhancing the existing covariate effect by reassigning $99\%$ of the older-subject cells to younger subjects.
Given this artificially large imbalance, the difference \eqref{eq:grifolswtdiff} for this modified cluster should be close to the inferred cluster size divided by the total number of cells.

\begin{figure}
    \includegraphics[width=\textwidth]{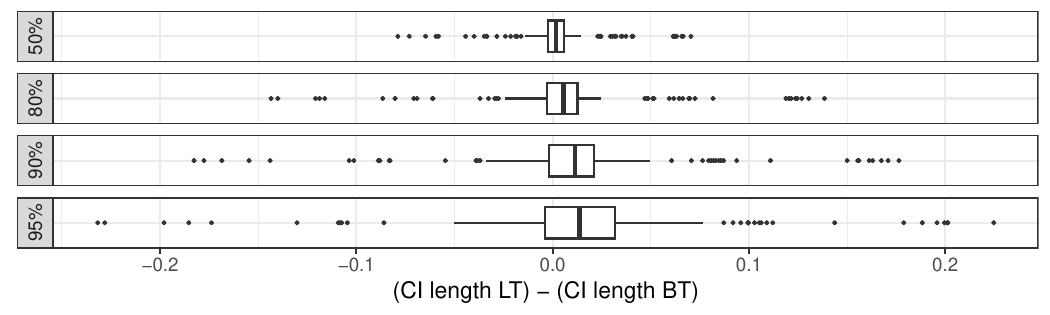}
    \caption{Each panel summarizes the distribution of 200 credible-interval length differences, where each difference is the difference between the credible-interval length for the lopsided-tree mixture model and that for the balanced-tree mixture model for some starting seed. Panel strips indicate credible-interval level.}
    \label{fig:pCIlengthdiff}
\end{figure}

Given the modified data, we fit a lopsided-tree model and a balanced-tree model for 200 different starting seeds and compare the behavior of the credible intervals of the difference \eqref{eq:grifolswtdiff} for this cluster. 
Rather than simply show all 400 credible intervals here (such plots are relegated to Figure~7 in the Supplemental Material), we summarize their behavior by first describing some desirable properties for them to have. 
One such property is whether it contains the truth, which for our purposes we set to be the cluster size. 
Also, an interval that contains the truth is more informative if it is short. 
In light of these properties, Table~\ref{tbl:GRIFOLSsim} shows the median interval length, the proportion of intervals that contain the truth, the proportion of lopsided-tree intervals that are a subset of the corresponding balanced-tree interval, and vice versa.
Here (and in Figure~\ref{fig:pCIlengthdiff}) we see that the lopsided-tree intervals tend to be longer than their balanced-tree counterparts.
Furthermore, there are many more starting seeds where the balanced-tree interval is a subset of the lopsided-tree interval than vice versa. 
We note that under some random seeds (14 seeds under the lopsided-tree model and 13 under the balanced-tree model), the posterior distribution fails to correctly identify the cluster and hence produces a tiny and incorrect credible interval centered at zero which (slightly) deflates the median interval length and proportion that contain the truth, but we find that this phenomenon influences the two different models to similar degrees and the general conclusions still hold even if these intervals are omitted. To avoid the suspicion of cherry picking, we did not exclude these seeds from our results.

\section{Discussion}
\label{sec:conc}

In addition to the comparisons made in this paper thus far between lopsided and balanced tree stick-breaking models, Section~4 of the Supplementary Materials presents yet another way the two tree structures behave differently from each other, this time regarding the phenomenon of component \textit{label switching} \citep[see][for an overview]{stephens2000dealing,Papastamoulis2016labelswitching}. 
Label switching has been observed to occur frequently in Markov chain Monte Carlo algorithms for mixture models that use  lopsided-tree stick-breaking models. 
Despite label switching being seen as a way to achieve convergence to the true posterior, such convergence can often become an unattainable goal to pursue.
The aforementioned section presents a simulation study where label switching occurs more often and to greater effect in a lopsided-tree mixture model than in a corresponding balanced-tree mixture model, and discusses reasons behind this behavior difference.

One limitation of the balanced tree model investigated in this work is that we assumed the tree is truncated at a fixed maximum depth. With a sufficiently large depth this causes little practical restriction in applications, but it does preclude theoretical analysis of such models as fully nonparametric processes with infinite-dimensional parameters. In particular, Pitman-Yor processes belong to the class of Gibbs-type priors \citep{gnedin2006exchangeable}, which have explicit forms for the predictive and clustering structure. Such an extension to balanced trees is of future interest. Another area of future work is to explore posterior-inference algorithms that avoid finite-dimensional approximations, allow covariates to be incorporated into mixture weights, and are computationally tractable.
A Pólya-urn sampler is one such method of avoiding finite-dimensional approximations, but it is not clear if such an approach could be extended to incorporate covariate dependence. 
On the other hand, \cite{Foti2012slice} introduces a slice sampler for the dependent Dirichlet process which perhaps could be modified for a dependent balanced-tree mixture model. 
Alternatively, we can still restrict $K$ to be finite but to also place a prior distribution on it as done in the mixture of finite mixtures e.g. \cite{gnedin2006exchangeable,de2013asymptotic,Miller2018,fruhwirth2021generalized}.

Another line of future work involves quantifying the prior dependence between random measures beyond the pairwise linear correlation used in Sections \ref{sec:priorcorrelation} and \ref{sec:corrGxGy}. Though our paper introduces dependence between random measures by stick breaking, much work has been done for the alternative approach of generating mixing weights using completely random measures \citep{kingman1967completely} in which dependence can be induced at the level of the underlying Poisson random measures. Under this framework, \cite{catalano2021measuring,catalano2023wasserstein} provide a general approach of quantifying the dependence between groups of random measures by measuring the dependence as distance from exchangeability using Wasserstein distance. If we place our generalized tree stick-breaking prior under the framework of completely random measures, it might be possible to glean additional insight into its dependence structure. 

\section*{Acknowledgements}

We gratefully acknowledge Scott White for assistance in processing flow cytometry data. 

This research was supported by the Translating Duke Health Initiative (Immunology) [AH, CC, LM], 
the Duke University Center for AIDS Research (CFAR), an NIH funded program (5P30 AI064518) [SW, CC],  
the National Institute of General Medical Sciences grant R01-GM135440 [AH, LM], 
and the National Science Foundation grant DMS-2013930 [LM].

\bibliographystyle{chicago}

\bibliography{mycites}

\newpage

\bigskip
\begin{center}
{\large\bf SUPPLEMENTARY MATERIAL}
\end{center}

\begin{description}

\item[Proofs:] Proofs of Theorems \ref{thm:meanvarcorr}, \ref{thm:summoment2},  \ref{thm:smooth}, \ref{thm:BTagree}, and Corollary~\ref{corollary:lowerbounds}. 

\item[Numerical experiment:] Numerical experiment regarding Corollary~\ref{corollary:lowerbounds}(c).

\item[Discussion:] Discussion on the computational cost of Gibbs step from the fixed-effects version of Algorithm~\ref{alg:gibbsrandomeffect}.

\item[Discussion:] Discussion on the impact of the tree on Markov chain Monte Carlo sampling regarding component label switching.

\item[Plot of 400 credible intervals:] Plot of the 400 credible intervals in the numerical study from Section~\ref{sec:influence}.

\end{description}

\section{Proofs}

\subsection{Proof of Theorem 1} 

\begin{proof}[Proof of Theorem 1]
This proof follows closely from Appendix 2 of \cite{Rodriguez2011}.
To prove the mean result (4), we have
\begin{align*}
\E \{G_{\xx}(A)\}
&= \E \left\{\sum_{k=1}^K W_{\xx,k} \delta_{\theta_{k}}(A)\right\}
= \sum_{k=1}^K \E (W_{\xx,k}) \text{pr}(\theta_{1} \in A)
= G_0(A).
\end{align*}

To prove the remaining results, let $X=G_{\xx}(A)$ and $Y=G_{\xx'}(A')$ and note 
\begin{align*}
\E (X Y)
&= \E \left[ \left\{\sum_{k=1}^K W_{\xx,k} \delta_{\theta_{k}}(A)\right\} \left\{\sum_{k=1}^K W_{\xx',k} \delta_{\theta_{k}}(A')\right\} \right] \\
&= \sum_{k=1}^K \E \left\{ W_{\xx,k} W_{\xx',k} \delta_{\theta_{k}}(A) \delta_{\theta_{k}}(A') \right\}
+ \sum_{k=1}^K \sum_{k' \neq k} \E \left\{ W_{\xx,k} W_{\xx',k'} \delta_{\theta_{k}}(A) \delta_{\theta_{k'}}(A')\right\} \\
&= G_0(A \cap A') a_{\xx,\xx'} + G_0(A)G_0(A') b_{\xx,\xx'}
\end{align*}
where $a_{\xx,\xx'} = \sum_{k=1}^K \E (W_{\xx,k} W_{\xx',k})$ and $b_{\xx,\xx'} = \sum_{k=1}^K \sum_{k' \neq k} \E (W_{\xx,k} W_{\xx',k'})$.
Because $a_{\xx,\xx'}+b_{\xx,\xx'}=1$, we have $\E(X) \E(Y) = G_0(A)G_0(A') (a_{\xx,\xx'}+b_{\xx,\xx'})$, and thus 
\begin{align*}
\Cov(X, Y) = 
\E (XY) - \E(X) \E(Y)
= \left\{G_0(A \cap A') - G_0(A)G_0(A')\right\} a_{\xx,\xx'}.
\end{align*}

The variance and covariances (5), (6), (7) result from appropriately setting $x=x'$ or $A=A'$ (or both).
The correlations (8) and (9) result from the definition $\corr(X,Y) = \{\E(XY) - \E(X) \E(Y)\} \{\Var(X) \Var(Y)\}^{-1/2}$, which requires the denominator to be nonzero.
\end{proof}

\subsection{Proof of Theorem 2}

\begin{proof}[Proof of Theorem 2]
For either tree, if $K=1$ the weight equals one a.s., which implies $a_{\x,\x'}=1$ as expressed in the theorem statement.
The remainder of the proof assumes that $K>1$, i.e., that at least one stick break occurs.

The lopsided-tree result comes mostly from Appendix~2 of \cite{Rodriguez2011}, which uses the incorrect assertion that the $K$th weight for a lopsided tree with $K$ leaf nodes is $W_K = V_K \prod_{k<K} (1-V_k)$, whereas in reality it is $W_K = \prod_{k<K} (1-V_k)$.
Once this correction is made, the corresponding result comes easily.

For the balanced-tree result, let $m = \log_2(K)$. 
For any $\bep \in B(\tau)$,
\begin{align*}
\E (W_{\xx,\bep} W_{\xx',\bep})
&= \prod_{l=1}^m \E ( V_{\xx,\ep_1 \cdots \ep_{l-1}} V_{\xx',\ep_1 \cdots \ep_{l-1}} )^{1-\ep_{l}} \E \{ (1-V_{\xx,\ep_1 \cdots \ep_{l-1}}) (1-V_{\xx',\ep_1 \cdots \ep_{l-1}}) \}^{\ep_{l}}.
\end{align*}
Because the distribution of any splitting variable does not depend on its node,
we have
\begin{align*}
\E (W_{\xx,\bep} W_{\xx',\bep}) =
    \prod_{l=1}^m c_{\xx,\xx'}^{1-\ep_{l}} d_{\xx,\xx'}^{\ep_{l}} = c_{\xx,\xx'}^{m-\sum_{l=1}^m \ep_l} d_{\xx,\xx'}^{\sum_{l=1}^m \ep_l}
\end{align*}
where 
$c_{\xx,\xx'}=\E \left( V_{\xx} V_{\xx'} \right)$ and $d_{\xx,\xx'}=\E \left\{ \left(1-V_{\xx}\right) \left(1-V_{\xx'}\right) \right\}$.
For any $k \in \{0, 1, \ldots, m\}$, the tree $\tau$ has exactly $\binom{m}{k}$-many leaves $\ep_1 \cdots \ep_m$ with the property $\sum_{l=1}^m \ep_l=k$ (i.e., leaves whose path from the root node chooses the ``right'' stick piece exactly $k$ times).
The desired sum $a_{\xx,\xx'}$ is then $\sum_{k=0}^m \binom{m}{k} c_{\xx,\xx'}^{m - k} d_{\xx,\xx'}^{k} = \left( c_{\xx,\xx'} + d_{\xx,\xx'} \right)^m$.
\end{proof}

\subsection{Proof of Theorem~3}

\begin{proof}[Proof of Theorem 3]
If both $\E (V_{\xx'}) \rightarrow \E (V_{\xx})$ and $\E (V_{\xx}V_{\xx'}) \rightarrow \E V_{\xx}^2$ as $\xx' \rightarrow \xx$,
then the proof of Theorem~2 gives us
\begin{align*}
\E (W_{\xx,\bep} W_{\xx',\bep})
&= [\E( V_{\xx} V_{\xx'})]^{m-\sum_{l=1}^m \ep_l} [\E \{ (1-V_{\xx}) (1-V_{\xx'}) \}]^{\sum_{l=1}^m \ep_l} \\
&\xrightarrow[\xx' \rightarrow \xx]{} [\E( V_{\xx}^2)]^{m-\sum_{l=1}^m \ep_l} [\E \{ (1-V_{\xx})^2 \}]^{\sum_{l=1}^m \ep_l} = \E (W_{\xx,\bep} W_{\xx,\bep}),
\end{align*}
which means $a_{\xx,\xx'} \xrightarrow[\xx' \rightarrow \xx]{} a_{\xx,\xx}$.
Thus, (9) $\xrightarrow[\xx' \rightarrow \xx]{} 1$.
\end{proof}

\subsection{Proof of Corollary 2.1}

\begin{proof}[Proof of Corollary 2.1]
We begin by proving the lopsided-tree lower bound.
For $K=1$ the lower bound is trivially satisfied.
Now suppose $K>1$ and define the function $g_K \colon [1/4,1/2] \rightarrow \mathbb{R}^+$ by $g_K(x) = x/(1-x) + x^{K-1}(1-2x)/(1-x)$. 
By the moment assumptions on the splitting variables, for any $\xx,\xx'\in\iset$ we have $\E(V_{\xx} V_{\xx'}) \in [1/4,1/2]$ and thus $a_{\xx,\xx'} = g_K(\E(V_{\xx} V_{\xx'}))$.
The derivative of $g_K$ is positive over its domain, which means $g_K$ has minimum $g_K(1/4) = 1/3 + (2/3)4^{1-K}$ and maximum $g_K(1/2)=1$.
Thus the cross-covariate correlation (9) for the lopsided tree is
\begin{align*}
(9) = \frac{a_{\xx,\xx'}}{\sqrt{a_{\xx,\xx} a_{\xx',\xx'}}} = \frac{g_K(\E(V_{\xx} V_{\xx'}))}{\sqrt{g_K(\E(V_{\xx}^2)) g_K(\E(V_{\xx'}^2))}} \geq \frac{1}{3} + \frac{2}{3} 4^{1-K}
\end{align*}
where the inequality follows from the stated bounds of each function $g_K$. 

Now we prove the balanced-tree lower bound.
Again we have $\E(V_{\xx} V_{\xx'}) \in [1/4,1/2]$ for any $\xx,\xx'\in\iset$. 
Then the cross-covariate correlation (9) for the balanced tree is
\begin{align*}
(9) &= \left[\frac{\E (V_{\xx} V_{\xx'})}{\{\E (V_{\xx}^2) \E (V_{\xx'}^2)\}^{1/2}}\right]^m
\geq \left[\frac{1/4}{\{1/2 \times 1/2\}^{1/2}}\right]^m = 2^{-m} = K^{-1}.
\end{align*}
If $\Var(V_{\xx})>0$ for all $\xx \in\iset$, then $\E(V_{\xx} V_{\xx'}) > 1/4$ for any $\xx,\xx'\in\iset$, which implies the lower bound in the preceding panel is strict.

Next we prove result~(b).
Because $\E(V_{\xx})=1/2$ for any $\xx$,
the correlation (9) is
\begin{align*}
(9) &= 
\left(\frac{\E (V_{\xx} V_{\xx'})}{ [\E (V_{\xx'}^2) \E (V_{\xx'}^2)]^{1/2} }\right)^m 
= \left(\frac{\corr(V_{\xx}, V_{\xx'}) \sqrt{\Var(V_{\xx}) \Var(V_{\xx'})} + 1/4}{ [\{\Var(V_{\xx})+1/4\} \{\Var(V_{\xx'})+1/4\}]^{1/2} }\right)^m.
\end{align*}
Because $\Var(V_{\xx})\Var(V_{\xx'})>0$ and $\corr(V_{\xx}, V_{\xx'}) < 1$ by assumption, the numerator inside the parentheses is positive and strictly smaller than $\sqrt{\Var(V_{\xx}) \Var(V_{\xx'})} + 1/4$, which itself is bounded above by the denominator inside the parentheses. 
Hence the entire fraction inside the parentheses is strictly between 0 and 1, which implies (9) shrinks to zero as $m \rightarrow \infty$.

Finally we prove result~(c).
By the representations of (9) for the two trees in the proof of part~(a), the ratio of (9) for the lopsided tree to (9) for the balanced tree is 
\begin{align*}
\frac{i_K(\E V_{\xx} V_{\xx'})}{\sqrt{i_K(\E V_{\xx}^2) i_K(\E V_{\xx'}^2)}},
\end{align*}
where we define the positive function $i_K \colon [1/4, 1/2] \rightarrow \mathbb{R}^+$ as $i_K(x) = g_K(x) / x^m$.
Once $\Var(V_{\xx})$, $\Var(V_{\xx'})$, and $\corr(V_{\xx}, V_{\xx'})$ are fixed, the preceding panel becomes a univariate function of $K$ and, if $K$ is allowed to be any real number, also is continuous in $K$ due to the function $i_K$ being continuous in $K$. 
Part~(a) of this corollary provides a positive lower bound for (9) for the lopsided tree whereas part~(b) of this corollary states that (9) for the balanced tree shrinks to zero as $K \rightarrow \infty$; these two parts along with the aforementioned continuity in $K$ together imply that the preceding panel grows unboundedly as $K \rightarrow \infty$. Thus the preceding panel is strictly larger than 1 when $K$ is sufficiently large.
\end{proof}

\subsection{Proof of Theorem 4} 

\begin{proof}[Proof of Theorem 4]
    We use the following formulation of the Wasserstein distance of any two probability measures $P$ and $Q$ over the same measurable metric space:
    \begin{align*}
        W(P,Q) = \sup_{f \in \mathscr{F}} \bigg\{ \int f \dif P - \int f \dif Q \bigg\}
    \end{align*}
    where $\mathscr{F}$ is the space of 1-Lipschitz functions $f$ defined on the metric space.
    Here the diameter of a subset $A$ of this metric space is the distance between the furthest two points in $A$, i.e., is $\text{diam}(A) = \sup_{x,y\in A} \rho(x,y)$ where $\rho$ is the space's metric. 

    Now let $P$ and $Q$ be the two tail-free processes assumed in the theorem statement. 
    Using any tree partition $\T_j$, the difference $\int f \dif P - \int f \dif Q$ can be decomposed into the sum $\sum_{A_{\bep} \in \T_j} [\int_{A_{\bep}} f \dif P - \int_{A_{\bep}} f \dif Q]$.
    Consider a single term in this sum. 
    For any 1-Lipschitz function $f$ and any subset $A_{\bep}$ from any $\T_j$, define $\bar{f}_{\bep} = \int_{A_{\bep}} f d\mu / \mu(A_{\bep})$ where $\mu$ is the Lebesgue measure. 
    Then the triangle inequality implies
    \begin{align*}
        \bigg| \int_{A_{\bep}} f \dif P - \int_{A_{\bep}} f \dif Q \bigg|
        &\leq \bigg| \int_{A_{\bep}} \bar{f}_{\bep} \dif P - \int_{A_{\bep}} \bar{f}_{\bep} \dif Q \bigg| + \bigg| \int_{A_{\bep}} (f-\bar{f}_{\bep}) \dif P \bigg| + \bigg| \int_{A_{\bep}} (f-\bar{f}_{\bep}) \dif Q \bigg| \\
        &\leq |\bar{f}_{\bep}| \big| P(A_{\bep}) - Q(A_{\bep}) \big| + \text{diam}(A_{\bep}) P(A_{\bep}) + \text{diam}(A_{\bep}) Q(A_{\bep}).
    \end{align*}
    If $|\bep| \leq M$, then $P(A_{\bep}) = Q(A_{\bep})$ by assumption, and so the preceding bound becomes $2 P(A_{\bep}) \text{diam}(A_{\bep})$. 
    If we decompose $\int f \dif P - \int f \dif Q$ using some $\T_j$ with $j \leq M$, we get 
    \begin{align*}
        W(P,Q) 
        = \sup_{f \in \mathcal{F}} \sum_{A_{\bep} \in \T_j} \bigg[\int_{A_{\bep}} f \dif P - \int_{A_{\bep}} f \dif Q\bigg] \leq 2 \sum_{A_{\bep} \in \T_j} P(A_{\bep}) \text{diam}(A_{\bep}). 
    \end{align*}
    With the assumption $\text{diam}(A_{\bep}) = 2^{-|\bep|}$, the upper bound becomes $2^{-j+1}$.
    Setting $j=M$ achieves the desired bound $2^{-M+1}$.
\end{proof}

\section{Numerical experiment regarding Corollary~2.1(c)}

Consider the ratio of the cross-covariate correlation (9) for the lopsided tree to that for the balanced tree as expressed in the proof of Corollary~2.1(c). Under the assumption of Corollary~2.1 that $\E (V_{\xx}) = 1/2$ for any $\xx$, this ratio becomes a function of four values: the number of leaves $K$, the correlation $\corr(V_{\xx}, V_{\xx'})$, and the two variances $\Var(V_{\xx})$ and $\Var(V_{\xx'})$. 
We have some intuition for how these four values impact the ratio. 
First, this ratio grows unboundedly as $K \rightarrow \infty$; this is asserted by the proof of Corollary~2.1(c).
Second, this ratio strictly decreases as the correlation $\corr(V_{\xx}, V_{\xx'})$ approaches~1; this is due to this correlation affecting this ratio only in the numerator $i_K(\E(V_{\xx} V_{\xx'}))$ and the function $i_K$ being strictly decreasing. 
(The impact of the two variances on the ratio is not as straightforward.) 

Using these observations, we conduct the following numerical experiment. 
We evaluate this ratio for $\corr(V_{\xx}, V_{\xx'}) = 0.99$, each $K \in \{4,8,16,32,64\}$, and many different values of $\Var(V_{\xx})$ and $\Var(V_{\xx'})$. Specifically, these variance values come from a $2501 \times 2501$ regular grid over the square $[0, 0.25]^2$. 
As shown in Table~\ref{table:minratiovalues}, we find for these $2501^2$ paired variance values and five $K$ values that (9) for the lopsided tree is strictly larger than that for the balanced tree.

\begin{table}
\begin{center}
\begin{tabular}{ |r|c|c|c|c|c| } 
 \hline
 number of leaves $K$ & 4 & 8 & 16 & 32 & 64 \\ \hline
 minimum ratio value & 1.000013 & 1.000022 & 1.000026 & 1.000029 & 1.000031 \\ 
 \hline
\end{tabular}
\end{center}
\caption{Mininum of $2501^2$ ratio values for various number of leaves.}
\label{table:minratiovalues}
\end{table}

\section{Computational cost of Gibbs step}
\label{sec:SM_gibbs_runtime}

Next we compare the computational cost of the Gibbs step in the algorithm below (which is a ``fixed-effects version'' of Algorithm~1 in the main text) between the two trees. 
For any iteration of the Gibbs sampler, each tree has the same number of internal nodes and hence their corresponding mixture models perform the same number of regressions.
From this viewpoint, the computational cost should be the same between the two stick-breaking schemes.

\begin{algorithm}[t]
\SetAlgoLined
\KwResult{Update each $R$-tuple $\bgamma_{\bep}$.}
\For{each internal node $\bep$ in binary tree $\tau$}{ 
    Let $D_{\bep}$ be the set of indices $i \in \{1,\ldots,n\}$ where $C_{\tau}(i)$ is a descendant of node $\bep$ \;
    \For{every $i \in D_{\bep}$} {
        Sample \PG data $[\omega_{i\bep} \mid \cdots] \sim \text{P\'olya-Gamma}(1, \, \bpsi(\xx_i)^\top \bgamma_{\bep})$ \;
    }
    Let $L_{\bep} \subset D_{\bep}$ be the set of indices corresponding to ``left descendants'' of node $\bep$ \;
    Update $\bgamma_{\bep}$ by drawing from full conditional
    \[[\bgamma_{\bep} \mid \cdots] \sim N_R(\bmu_{\bgamma_{\bep}}, \bSigma_{\bgamma_{\bep}})\]
    where, letting $\bkappa_{\bep} = (1_{i \in L_{\bep}} - 0.5)_{i \in D_{\bep}}$
    and $\X_{\bep}$ be the $|D_{\bep}| \times R$ matrix with row entries $\bpsi(\xx_i)^\top$ for only those $i \in D_{\bep}$,
    \begin{align*}
        \bmu_{\bgamma_{\bep}} = \bSigma_{\bgamma_{\bep}} \left[\X_{\bep}^\top \bkappa_{\bep} + \bSigma_{\bgamma}^{-1} \bmu_{\bgamma} \right], \quad
        \bSigma_{\bgamma_{\bep}}^{-1} = \X_{\bep}^\top \text{diag}(\omega_{1{\bep}}, \ldots, \omega_{|D_{\bep}| \bep}) \X_{\bep} + \bSigma_{\bgamma}^{-1}.
    \end{align*}
}
\caption{Gibbs step to update logit-normal coefficient under any binary tree.}
\label{alg:gibbs}
\end{algorithm}

\begin{figure}
\begin{center}
\includegraphics[width=\textwidth]{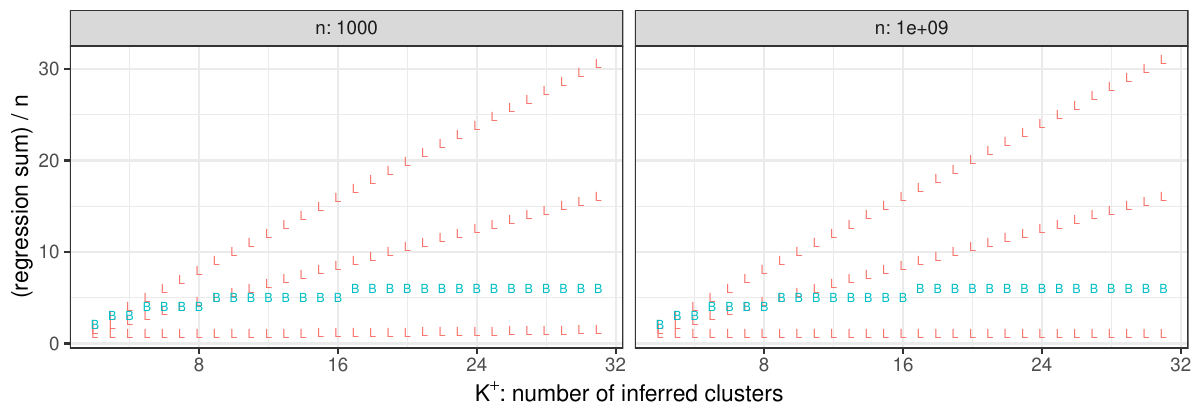}
\end{center}
\caption{Scaled sum values of $(1 + n^{-1} K^+ (K^+ - 1)/2)$, $n^{-1} (K^+ (n - K^+ + 1) + K^+ (K^+ - 1)/2)$, $(K^+ + 1)/2$, and $(1 + \lceil \log_2 K^+ \rceil)$ for various combinations of $n$ and $K^+$. The first three expressions are displayed as `L' while the fourth expression is displayed as `B'.}
\label{fig:regsumgibbs}
\end{figure}

However, we find that training a balanced-tree mixture model takes less time than training its lopsided-tree counterpart for the GRIFOLS flow-cytometry and simulated data sets in the main text.
To see why, for a tree $\tau$ and leaf-assignment $C_{\tau}$, the computational cost of the Gibbs step can be roughly identified with the sum (over all $i=1,\ldots,n$) of the number of $C_{\tau}$'s regressions involving observation $i$, which itself equals the sum (over all $i=1,\ldots,n$) of the number of ancestors of leaf node $C_{\tau}(i)$.
The sum for a balanced tree equals $n \log_2 K$ because all $n$ observations have $\log_2 K$ ancestors.
The sum for a lopsided tree, however, decreases as observations are allocated to leaf nodes closer to root (meaning the observations are involved in fewer regressions) and increases as observations are allocated further from root.
We use this reasoning to get the sum's extrema for a lopsided tree.
If $K^+$ clusters are inferred, the sum is smallest when the leaf-node child of the root contains $(n-K^++1)$ observations and $(K^+-1)$ other leaf nodes each contain just one observation, which results in the sum equaling
\begin{equation*}
    (n - K^+ + 1) + (2 + 3 + 4 + \cdots + K^+) = n + K^+ (K^+ - 1)/2.
\end{equation*}
Similarly, the sum is largest if the $(n - K^+ + 1)$ observations are contained in the nonempty leaf node \textit{furthest} from root, which results in the sum equaling
\begin{equation*}
    K^+ (n - K^+ + 1) + (1 + 2 + \cdots + (K^+-1)) = K^+ (n - K^+ + 1) + K^+ (K^+ - 1)/2.
\end{equation*}
If $n$ is fixed and much larger than $K^+$, the sum's minimum divided by $n$ is roughly constant in $K^+$ whereas the sum's maximum divided by $n$ is roughly linear in $K^+$, as shown in Figure~\ref{fig:regsumgibbs}.

Though these extreme scenarios provide useful computational cost bounds for a lopsided tree, seldom will they be even remotely close to how a leaf assignment $C_{\tau}$ will allocate observations.
On the other hand, what exactly is a more ``typical'' allocation, and can we find an analytical representation for it or for a reasonable approximation of it?
The answer to the former question (and hence also the latter question) depends on, among other things, the size of the data's clusters and how well the mixture model infers these sizes.
For simplicity, in the following exposition we will assume the mixture model infers $K^+$ leaf nodes each containing $n/K^+$ observations, in which case the sum for a lopsided tree becomes $n (K^+ + 1)/2$, which again is linear in $K^+$.

The final hurdle for this computational cost comparison between a lopsided tree and a balanced tree is that our three sum quantities for a lopsided tree, $n + K^+ (K^+ - 1)/2$, $K^+ (n - K^+ + 1) + K^+ (K^+ - 1)/2$, and $n (K^+ + 1)/2$, are functions of $n$ and $K^+$ while the sum quantity for a balanced tree, $n \log_2 K$, is a function of $n$ and $K$.
How can we relate $K$ to $K^+$?
We might optimistically assume the practitioner chose $K$ to be the smallest power of two greater than or equal to $K^+$,
but we will ``leave room for error'' by instead assuming $K$ is the smallest power of two greater than or equal to $2 K^+$,
which implies $\log_2 K = (1 + \lceil \log_2 K^+ \rceil)$.
Now all four sum quantities can be expressed as functions of $n$ and $K^+$ as shown in Figure~\ref{fig:regsumgibbs}.
The (scaled) sum for a balanced tree grows logarithmically in $K^+$ whereas the (scaled) sum for the ``typical'' lopsided-tree scenario grows linearly, which explains why we find that training a balanced-tree mixture is usually faster than training its lopsided-tree counterpart.

\section{Impact of the tree on Markov chain Monte Carlo sampling}
\label{sec:labelswitch}

Label switching has been observed to occur frequently in Markov chain Monte Carlo algorithms for mixture models that use standard, or lopsided-tree, stick-breaking models. 
Despite label switching being seen as a way to achieve convergence to the true posterior, such convergence can often become an unattainable goal to pursue.
In this section we design a simulation experiment 
to compare the frequency of label switching under stick-breaking models with different tree structures. In particular, our results indicate that label switching can occur more often and to greater effect in the trained lopsided-tree mixture model than in the trained balanced-tree mixture model. We also discuss reasons behind this phenomenon.

\begin{figure}
    \includegraphics[width=\textwidth]{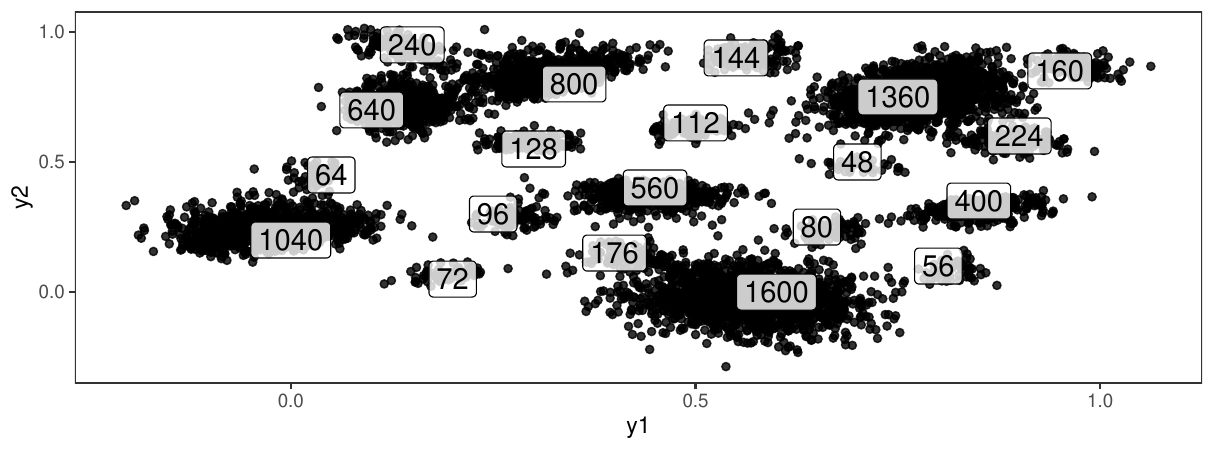}
\caption{Data generated by 20 skew-normal distributions. The boxed numbers are the number of observations generated from each distribution.}
\label{fig:datdecay_sm}
\end{figure}

We generate $n=\num{8000}$ observations each from one of 20 skew-normal distributions shown in Figure~\ref{fig:datdecay_sm}; these cluster counts are listed as the $20$-tuple 
\begin{align*}
    8 \cdot (200, 170, 130, 100, 80, 70, 50, 30, 28, 22, 20, 18, 16, 14, 12, 10, 9, 8, 7, 6).
\end{align*}
For each cluster, the observations are evenly assigned one of the eight covariates in the set $\{1\} \times \{0,1\}^3$.
To the data we fit a lopsided-tree mixture model and a balanced-tree mixture model using the Gibbs step in Algorithm~\ref{alg:gibbs} with $K=32$, prior mean $\bmu_{\gamma} = 0_4$, and prior covariance $\bSigma_{\gamma} = 10 I_4$.
Each chain burns in \num{100000} steps before sampling every $100$ steps to keep \num{10000} posterior draws.

\begin{figure}
    \includegraphics[width=\textwidth]{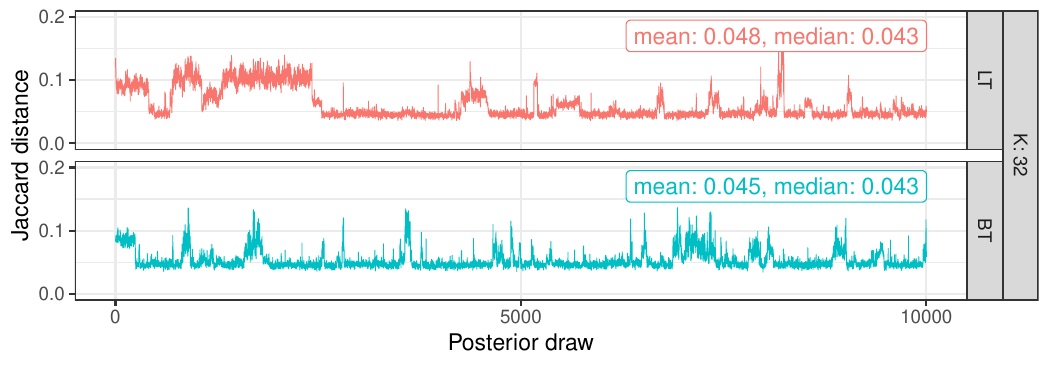}
    \caption{Jaccard distances at each posterior draw.}
    \label{fig:20clu_jm}
\end{figure}

Figure~\ref{fig:20clu_jm} shows the Jaccard distance between each chain's clustering at the $j$th posterior draw (for all $j = 1,\ldots,\num{10000}$) and the true clustering \citep{Jaccard1912}.
This distance provide a sense of a chain's clustering performance \textit{and} its mixing behavior.
The two models have roughly the same \textit{median} Jaccard distance of $0.043$, but the \textit{mean} Jaccard distance is slightly larger for the lopsided-tree mixture model ($0.048$ vs $0.045$).
The lopsided-tree chain also seems to get stuck in local modes for longer than the balanced-tree chain does.

\begin{figure}
     \centering
     \begin{subfigure}[b]{0.47\textwidth}
         \centering
         \includegraphics[width=\textwidth]{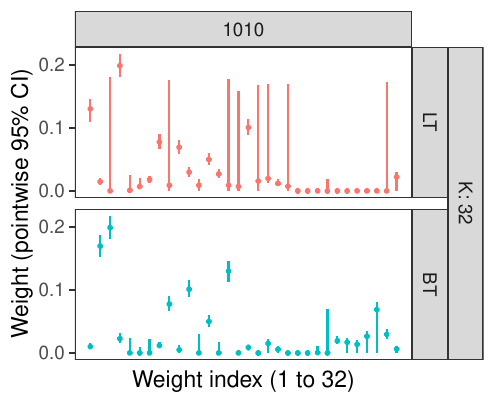}
         \caption{Pointwise $95\%$ credible intervals of weights inferred by models trained on the two data sets.}
         \label{fig:1010wtCI}
     \end{subfigure}
     \hfill
     \begin{subfigure}[b]{0.47\textwidth}
         \centering
         \includegraphics[width=\textwidth]{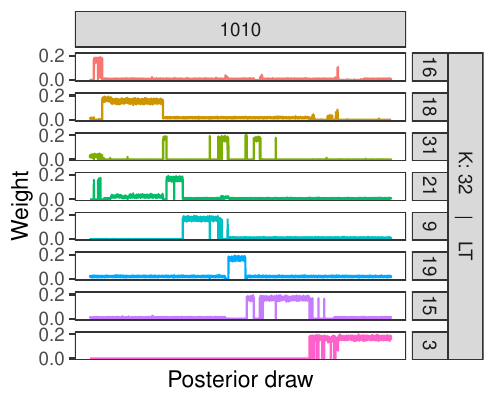}
         \caption{Weights inferred by the lopsided-tree mixture for indices corresponding to wide credible intervals.}
         \label{fig:1010wtLT}
     \end{subfigure}
     \hfill
    \caption{Selected mixture-weight inference for the covariate vector $1010$.}
    \label{fig:1010wt}
\end{figure}

\begin{figure}
    \begin{subfigure}{\textwidth}
    \includegraphics[width=\textwidth]{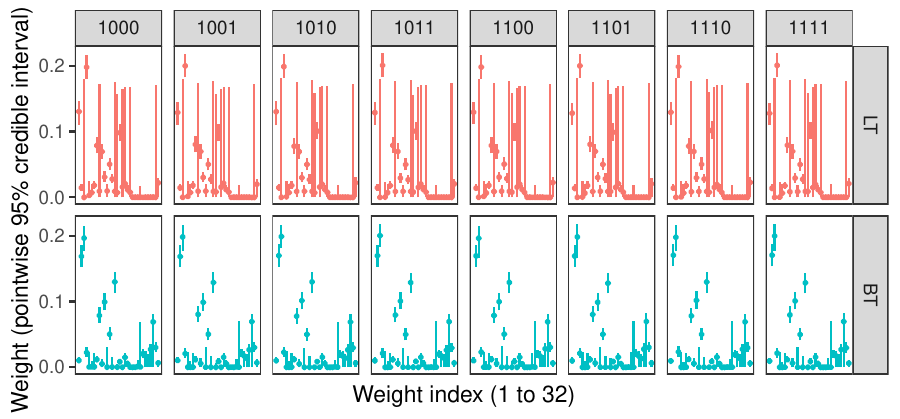}
    \end{subfigure}
    \caption{Inferred pointwise $95\%$ credible intervals for various covariate combinations.}
    \label{fig:ptwise}
\end{figure}

What accounts for this difference in clustering behavior?
Figure~\ref{fig:1010wt} shows selected mixture-weight inference for the covariate combination $1010$.
The results are similar for the other covariate values as shown in Figure~\ref{fig:ptwise}.
In Figure~\ref{fig:1010wtCI}'s lopsided-tree panel, no credible interval is \textit{centered} around the second largest weight value of $0.17$, which might foster the belief that this lopsided-tree mixture model fails to capture this weight, but the \textit{heights} of the eight curiously wide $95\%$ credible intervals are roughly that same value. 
This observation together with Figure~\ref{fig:1010wtLT} imply the eight wide lopsided-tree credible intervals are a result of label switching in the Markov chain. 
Similarly, Figure~\ref{fig:1010wtBT_sm} in the Supplemental Material implies the wide credible intervals in Figure~\ref{fig:1010wtCI}'s balanced-tree panel are also a result of label switching. 
Indeed, Figure~\ref{fig:1010wtCI_sorted_sm} in the Supplemental Material shows these wide credible intervals disappear and the data's weights are captured if we ``resolve'' the label switching by performing the following post-hoc sorting of weights: for each posterior draw $j$ we rearrange (in decreasing order) the elements of the chain's corresponding weight vector $w_{1010}^{(j)}$, then for each newly rearranged weight index $\tilde{k}$ we compute credible intervals for the values in the set $ \{w_{1010, \tilde{k}}^{(j)} : j = 1,\ldots,\num{10000}\}$.

Although this post-hoc sorting seems to resolve the wide credible intervals for \textit{this} example, the procedure requires several assumptions to work and hence cannot be applied generally. 
Until a general solution is created, it seems prudent to ``nip the problem in the bud'' by mitigating the impact of label switching on inference.
But why does label switching seem to affect the lopsided-tree mixture's inference more than balanced-tree mixture's? We conjecture that the behavior difference stems from two tree-related mechanisms. 

First, consider how much friction there is to label switch between two leaf nodes for either tree structure. 
The amount of friction seems to increase with the number of splitting variables affected by the proposed switch. 
If the two leaves are graphically far apart, the proposed switch would affect many splitting variables and hence be unlikely to occur. 
Under this reasoning, switching is most likely to occur between two leaves graphically near each other. 
In a balanced tree, this means two leaves that share the same parent.
In a lopsided tree, this means two leaves separated by at most one tree level.
For conceptual simplicity, we consider only such pairs and call these leaves \textit{adjacent} to each other.
Because a lopsided tree has $K-1$ adjacent pairs whereas a balanced tree has $K/2$, a lopsided tree has more adjacent pairs from which switching can occur. 

Second, a label switch between two adjacent lopsided-tree leaves can create a ``chain reaction'' of label switches either up or down the entire tree. 
If a switch occurs between adjacent lopsided-tree leaf-pair $(\ep 0, \ep 10)$ for some internal node $\ep$, the adjacent pair $(\ep 10, \ep 110)$ becomes ``newly eligible'' for switching, and if \textit{that} switch eventually occurs, the adjacent pair $(\ep 110, \ep 1110)$ becomes ``newly eligible'' for switching, and so on, which creates the possibility of a chain reaction in the direction of increasing tree level (though a chain reaction can just as easily occur in the other tree direction).
This might explain the behavior in Figure~\ref{fig:1010wtLT}, where the $0.17$ mass moves almost sequentially through eight nodes. 
In contrast, because any balanced-tree leaf is adjacent to exactly one other leaf and because adjacency is a symmetric binary relation, any switch between two adjacent balanced-tree leaves will likely stay localized i.e. such a switch will likely not create any ``newly eligible'' pairs.
This balanced-tree conjecture is supported by the chain behavior in Figure~\ref{fig:1010wtBT_sm}. 

These two proposed mechanisms seem to explain the observed Markov chain Monte Carlo behavior difference between a lopsided tree and a balanced tree. 
Overall, our aforementioned inference observations indicate that if we consider the parameter space $\Theta$ and posterior local modes corresponding to label permutations, the regions between these local modes appear to be deeper valleys for the balanced-tree posterior than they are for the lopsided-tree posterior, making it more difficult to jump between different labelings for a balanced tree than for a lopsided tree.

\begin{figure}
     \centering
     \begin{subfigure}[b]{\textwidth}
         \centering
         \includegraphics[width=\textwidth]{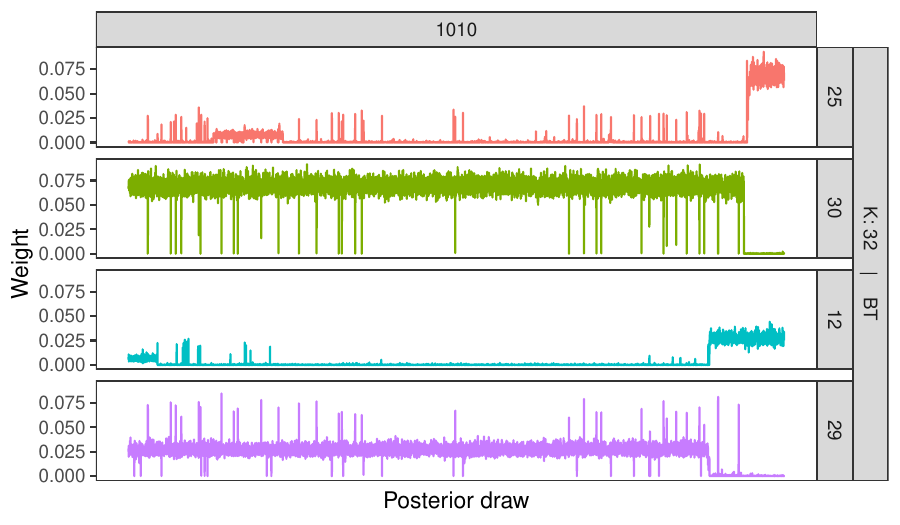}
         \caption{Weights inferred by the balanced-tree mixture model for indices corresponding to wide credible intervals.}
         \label{fig:1010wtBT_sm}
     \end{subfigure}
     ~
     \begin{subfigure}[b]{0.47\textwidth}
         \centering
         \includegraphics[width=\textwidth]{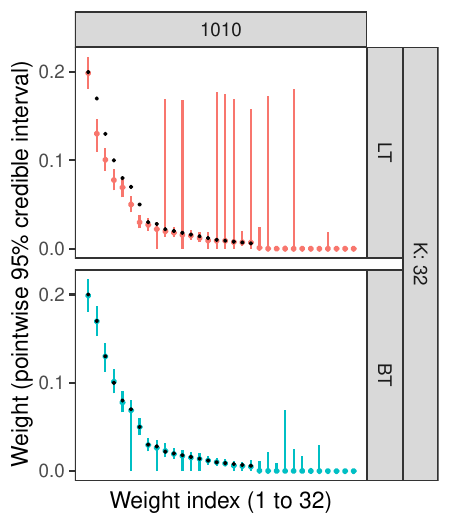}
         \caption{Inferred pointwise $95\%$ credible intervals that are sorted by median weight. Points indicate the data's true weight values.}
         \label{fig:1010wtCIr_sm}
     \end{subfigure}
     \hfill
     \begin{subfigure}[b]{0.47\textwidth}
         \centering
         \includegraphics[width=\textwidth]{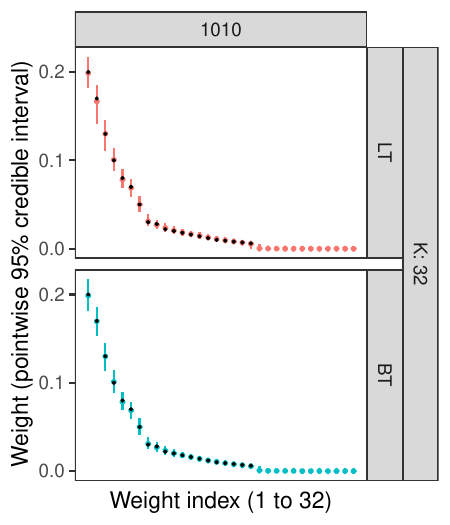}
         \caption{Inferred weights that are sorted before taking pointwise $95\%$ credible intervals. Points indicate the data's true weight values.}
         \label{fig:1010wtCI_sorted_sm}
     \end{subfigure}
     \hfill
    \caption{Selected mixture-weight inference for $\xx = 1010$.}
    \label{fig:1010wt_sm}
\end{figure}

\begin{figure}[p]
    \includegraphics[width=\textwidth]{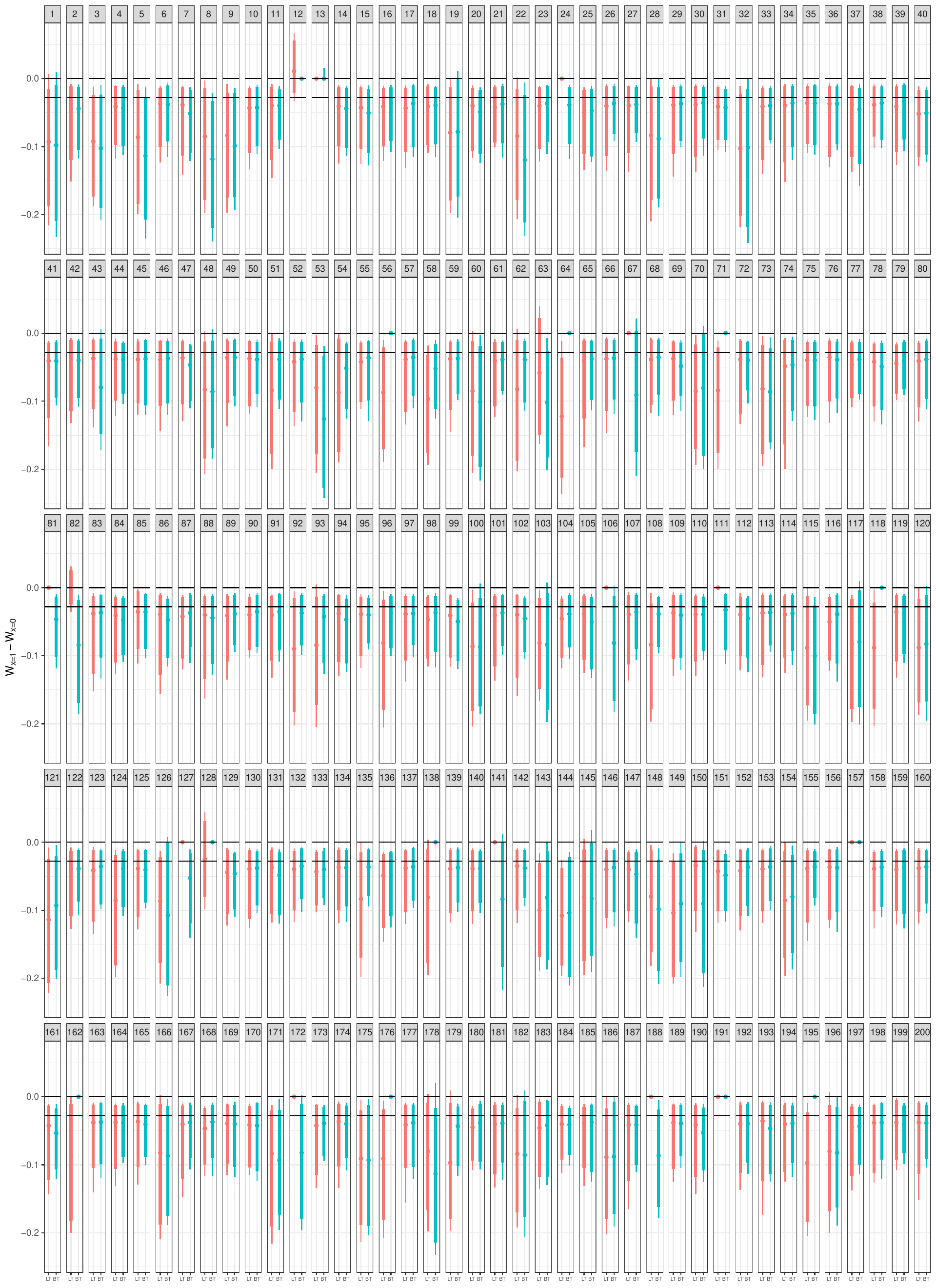}
\caption{Credible intervals of the 200 lopsided-tree models and 200 balanced-tree models trained on the modified data described in Section 4.5 of the main text.}
\label{fig:GRIFOLSnumstudyCI}
\end{figure}

\end{document}